\documentclass[journal,twoside]{IEEEtran}

\ifCLASSOPTIONcompsoc
  \usepackage[nocompress]{cite}
\else
  \usepackage{cite}
\fi

\ifCLASSINFOpdf
\else
\fi

\usepackage{algpseudocode}
\usepackage{algorithm}
\usepackage{amsmath,amssymb,amsthm,amsfonts}
\usepackage{breqn}
\usepackage{mathtools}
\usepackage{cases}
\usepackage{mathrsfs}
\usepackage{accents}
\usepackage{enumitem}
\usepackage{tabu}
\usepackage{stfloats}
\usepackage{mathptmx}
\usepackage{makecell}
\usepackage{caption}
\usepackage{graphicx}
\usepackage{epstopdf}

\usepackage{times}
\newcommand{\subparagraph}{}
\usepackage{titlesec}
\titlespacing\section{0pt}{4pt plus 4pt minus 2pt}{4pt plus 2pt minus 2pt}
\titlespacing\subsection{0pt}{3pt plus 2pt minus 2pt}{3pt plus 2pt minus 2pt}
\usepackage{array}
\usepackage{subfig}
\usepackage[export]{adjustbox}
\usepackage{xcolor}
\usepackage[T1]{fontenc}
\usepackage[utf8]{inputenc}
\usepackage{url}
\usepackage[normalem]{ulem}
\useunder{\uline}{\ul}{}
\usepackage{balance}
\usepackage{epstopdf}
\usepackage{scrextend}
\usepackage{multirow}
\let\OLDthebibliography\thebibliography
\renewcommand\thebibliography[1]{
  \OLDthebibliography{#1}
  \setlength{\parskip}{0.3pt}
  \setlength{\itemsep}{0.5pt plus 0.6ex}
}
\newlength{\dhatheight}

\newcommand{\etal}{\textit{et al.}}
\newcommand{\Fig}{Fig.}
\newcommand{\Eq}{Eq.}
\newcommand{\Figs}{Figs.}

\newcolumntype{P}[1]{>{\centering\arraybackslash}p{#1}}
\newcolumntype{M}[1]{>{\centering\arraybackslash}m{#1}}
\makeatletter
\newcommand*\bigcdot{\mathpalette\bigcdot@{1}}
\newcommand*\bigcdot@[2]{\mathbin{\vcenter{\hbox{\scalebox{#2}{$\m@th#1\bullet$}}}}}
\makeatother

\begin{document}
\title{Understanding the Dynamics of Drivers' Locations for Passengers Pickup Performance: A Case Study}

\author{Punit~Rathore,~Ali~Zonoozi,~Omid~Geramifard,~Tan~Kian~Lee 
\IEEEcompsocitemizethanks{\IEEEcompsocthanksitem  Punit Rathore is with the Senseable City Lab, Department of Urban Planning and Studies, Massachusetts Institute of Technology, Cambridge, USA.
\protect
E-mail: prathore@mit.edu
\IEEEcompsocthanksitem Ali Zonoozi is with the GrabTaxi  Holdings Ltd, Singapore.
\protect
E-mail: ali.zonoozi@grab.com.
\IEEEcompsocthanksitem Omid Geramifard is with the  Air Asia Group, Singapore.
\protect
E-mail: omid.geramifard@gmail.com.
\IEEEcompsocthanksitem  Tan Kian Lee is with the School of Computing, National University of Singapore, Singapore.
\protect
E-mail: tankl@comp.nus.edu.sg
}
}

\IEEEtitleabstractindextext{%
\begin{abstract}
With the emergence of e-hailing taxi services, a growing number of scholars have attempted to analyze the taxi trips data to gain insights from drivers' and passengers' flow patterns and understand different dynamics of urban public transportation. Existing studies are limited to passengers' location analysis e.g., pick-up and drop-off points, in the context of maximizing the profits or better managing the resources for service providers. Moreover, taxi drivers' locations at the time of pick-up requests and their pickup performance in the spatial-temporal domain have not been explored. In this paper, we analyze drivers' and passengers' locations at the time of booking request in the context of drivers' pick-up performances. To facilitate our analysis, we implement a modified and extended version of a co-clustering technique, called sco-iVAT, to obtain useful clusters and co-clusters from big relational data, derived from booking records of Grab ride-hailing service in Singapore. We also explored the possibility of predicting timely pickup for a given booking request, without using entire trajectories data. Finally, we devised two scoring mechanisms to compute pickup performance score for all driver candidates for a booking request. These scores could be integrated into a booking assignment model to prioritize top-performing drivers for passenger pickups.
\end{abstract}

\begin{IEEEkeywords}
ride-hailing service, e-hailing, driver locations, pick-up locations, big relational data, co-clustering.
\end{IEEEkeywords}}

\maketitle

\IEEEdisplaynontitleabstractindextext

\IEEEpeerreviewmaketitle


\section{Introduction}\label{sec:introduction}
The emergence of the online ride-hailing platforms like Grab, Uber, Lyft etc., have made a revolution in the industry of the mobility service~\cite{conway2018trends}. These e-hailing services offer convenient passenger-to-taxi booking services and improve the quality of urban taxi services. The prevalence of smartphones and sensors-equipped vehicles enabled online ride-hailing platforms to gather large amounts of data and analyze them to provide both drivers and passengers a better experience~\cite{su2019review}.

In a typical ride-hailing scenario, a passenger places a booking request through a smartphone app, which is then broadcast to many drivers, and fulfilled by dispatching the most suitable, nearby driver available to serve the ride~\cite{maciejewski2016assignment,ozkan2020dynamic}. For each booking request, the service provider searches and tracks 
available drivers within a range near a passenger's location. Generally, the nearest driver from a passenger gets the request first, and if the nearest driver does not accept, this ping is sent to the next nearest driver and so on, until the ping is accepted by a driver available within a range of the requesting passenger. Many service providers also consider travel time or estimated time of arrival for pickup to determine the closest driver on the road network. \textit{Estimated time of arrival} (ETA) is usually derived from the historical trips data of a road network. Some service providers also take into account the acceptance rate, cancellation rate, and other features to match a driver to a passenger.

\begin{figure}
\centering
\includegraphics[width=0.35\textwidth]{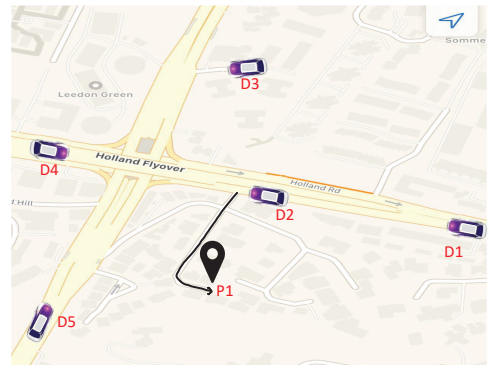}
\caption{A driver-allocation scenario for a passenger pickup.}%
\label{Fig:DriverAllocation}
\end{figure}

Locating the driver based on the distance or ETA may not always be the best choice for a passenger pickup.  \Fig~\ref{Fig:DriverAllocation} shows such an example where driver $D2$ would be assigned to the passenger $P1$ based on the lowest ETA for passenger pickup. However, since driver $D2$ is on the highway, he/she might miss the next intersection or suggested route/turn (by the service provider) due to high speed. Consequently, driver $D2$'s \textit{actual time of arrival} (ATA) may become higher than the ETA of other nearby drivers. In this example, driver $D1$, whose ETA was higher than the driver $D2$'s ETA, might be a better choice for passenger pickup as he/she would get enough time to follow the suggested/shortest route and arrive at pickup location earlier than $D1$'s ATA.

\begin{figure}
\centering
\includegraphics[width=0.4\textwidth]{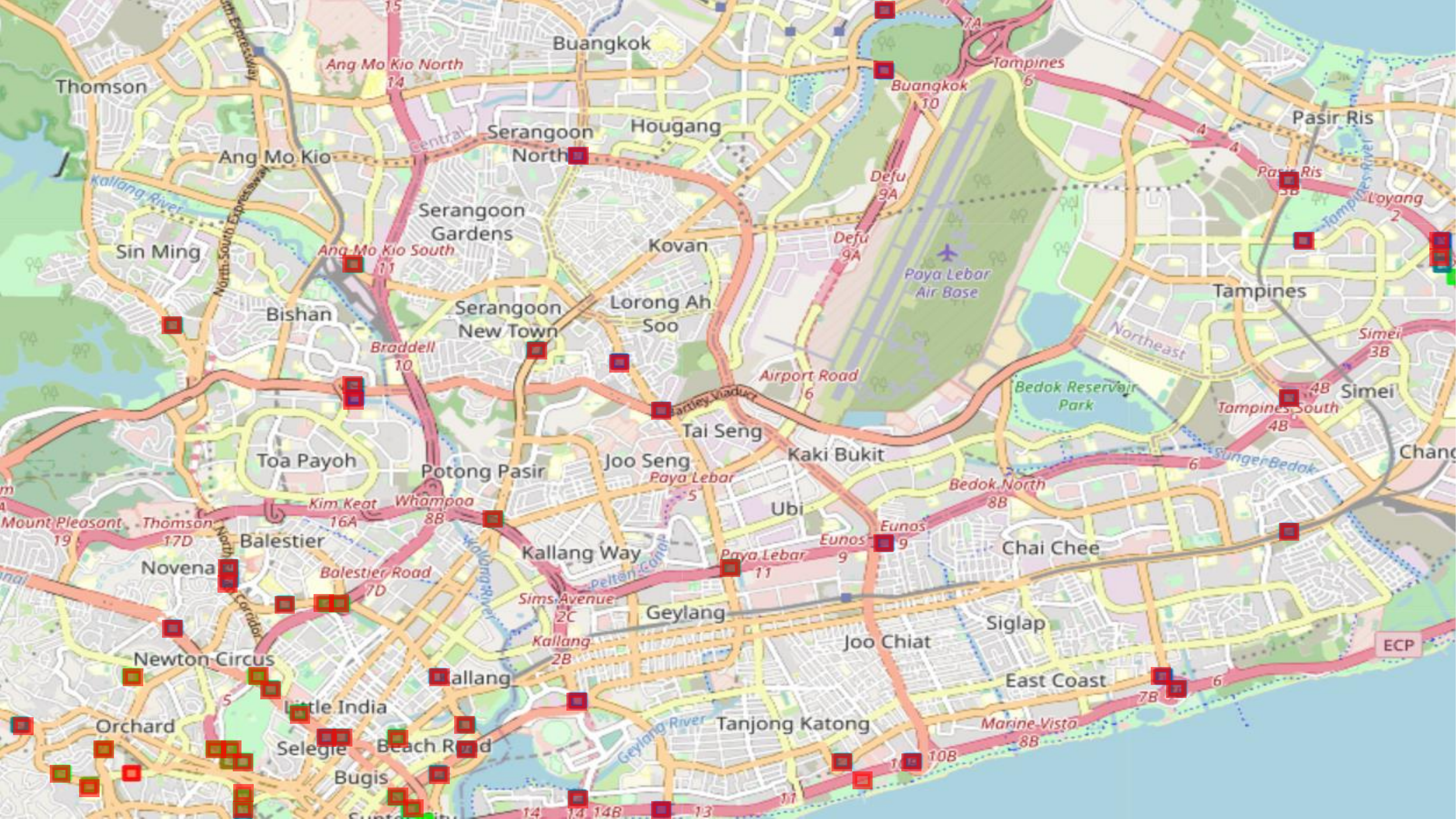}
\caption{\textit{Driver location} grids ($152m \times 152m$) with average speed greater than $35$ km/h and late pickup performance.}
\label{Fig:GeohashesHighway}
\end{figure}

In our preliminary analysis on a booking data (refer to Section~\ref{sec:methodology}), we extracted all the grids (drivers' locations at the time of booking request) on highways (speed > $35$kmph) of Singapore road network where drivers took longer ($>5 mins)$ than ETA for passenger pickup in at least $ 100$ bookings.~\Fig~\ref{Fig:GeohashesHighway} shows these grids with red squares where many grids lie near the intersections on highways, supporting our conjecture for late pickup. While there may be other factors behind late pick-up corresponding to these drivers' locations, identification and characterization of such locations may help to choose the best driver for passenger pickup. 

\setlength{\textfloatsep}{-0.1pt}

There are some studies~\cite{su2018understanding,su2019review,goel2016optimal,niu2019predicting,qian2020understanding} that analyzed instant booking data to identify the relevant pickup and drop-off hotspots in the spatial-temporal domain and studied the relationships between pickup and drop-off locations. Such studies may help drivers to increase their profits and minimize their travel distance by suggesting these hotspots. However, most of these studies focused only on pickup and drop-off location data analysis due to the absence of driver locations when a booking is requested by a passenger. Taxi drivers' locations at the time of booking request and their effect on pickup performance have not been explored yet in the literature.

In this article, we analyze drivers' and passengers' locations during pickup request and other important relevant features e.g., ETA, ATA etc., to obtain driver- and location-specific insights that may be useful to determine the best driver for a passenger pickup. We use one-month booking records of Grab ride-hailing service~\cite{huang2019grab} in Singapore. We denote the driver's location at the time of booking request with "\textit{driver location}" and passenger's location with "\textit{pickup location}". 

In this paper, we aim to answer the following specific questions about the dynamics of drivers and their locations that may influence passenger pickup performance.

\begin{itemize}
    \item Are there any spatial pattern on \textit{driver locations} that may influence ETA for passenger pickup? For example, there may be some \textit{drivers' locations} for which most drivers are always late or always on time for pickup.
    \item  Are there any driver-specific spatial patterns influencing passenger pick-ups? There may be some locations where specific drivers are mostly late or mostly on time for passenger pickups.
    \item  Are there any location-specific patterns among drivers for passenger pick-ups? For example, there may be some drivers who are mostly late at specific locations for passenger pickups. 
    \item Are there any pattern among drivers for pick-ups? For example, there may be some drivers who are mostly late or mostly on time for passenger pickups.
\end{itemize}

The answers to these questions can assist us to characterize such drivers and locations which may be helpful to determine the best driver for pickup. For example, such drivers may be characterized based on age, their driving experience,  familiarity, performance at a given location,  booking acceptance/cancellation rates, and the number of bookings accepted in a day (to assess tiredness) etc. Similarly, locations can be characterized by traffic density, location type (e.g. intersections, highway, downtown etc.),  and total bookings made at that location in temporal domain etc. These characteristics can be used to score the drivers (overall and location-wise) and include them in driver allocation framework to prioritize/deprioritize certain drivers for a booking assignment.

To answer the above questions, we need to analyze how drivers and their locations are clustered in the spatio-temporal domain. Although each of these (four) questions can be answered separately by individual clustering of drivers and locations based on the pickup performance, however, in our work, we answer all of them using a single technique employing co-clustering or bi-clustering. Simultaneous clustering of the columns and rows of the data matrix, also known as co-clustering~\cite{dhillon2003information}, identifies a subset of rows that are similar across a subset of columns in a rectangular relational data. Many applications, such as gene-expression data in biology application~\cite{cho2004minimum}, word-document data in text mining, and user-rating data in recommendation systems, use co-clustering to analyze the relational data. 

In this work, we implement a modified and extended version of a co-clustering technique sco-VAT~\cite{park2009visualization}, which will call sco-iVAT (\textit{scalable co-clustering using improved visual assessment of tendency}), and apply it to a big driver-locations relational data to identify (i) subset(s) of drivers that are similar (ii) subset(s) of locations that are similar, and  (iii) subset(s) of drivers that are similar across a subset(s) of locations, for passenger pickup performance. In addition to exploratory analysis, we also explore the possibility of predicting timely pickup of a passenger for a given booking request, without using entire GPS traces of drivers. We also study the importance of different predictors for timely-pickup classification. Eventually, we devise two driver scoring mechanisms based on drivers' overall and location-wise pickup performance and features, which can be used to complement existing driver allocation (dispatch) model. To our knowledge, this is the first study of its kind analyzing drivers' location at the time of booking request for passenger pickup performance.

Section~\ref{sec:relatedworks} reviews the relevant studies and Section~\ref{sec:preliminary} presents a preliminary on techniques used in this work. Our developed co-clustering algorithm, sco-iVAT, is discussed in Section~\ref{sec:sco-iVAT}. Sections~\ref{sec:methodology}-
\ref{sec:timelypickup} discuss the dataset, methodology, and results of our analysis. Sections~\ref{sec:scoring} introduces two scoring mechanism for driver pickup performance, followed by conclusions in Section~\ref{sec:conclusion}.


\section{Related Work}\label{sec:relatedworks}

Several studies~\cite{su2019review} have mined taxi trips and booking data provided by ride-hailing services and taxi cabs in the context of fleet management~\cite{oda2018movi}, taxi demand prediction~\cite{xu2017real,niu2019predicting}, dynamic pricing~\cite{qian2017time} 
, distributed ride sharing~\cite{bathla2018real}, optimal pick-up point selection for ride-sharing~\cite{goel2016optimal}, driver-passenger matching~\cite{maciejewski2016assignment}, taxi recommendation~\cite{lee2008analysis}, next pickup-point prediction~\cite{veloso2011urban,hwang2015effective}, and identifying patterns and dynamics of behaviours or activities and spatial interactions~\cite{liu2010uncovering,zhan2016graph,jiang2017multi,hu2018taxi,qian2020understanding}. 

Xu~\etal~\cite{xu2017real} developed a sequential learning-based taxi demand prediction model by learning historical demand patterns from the yellow and green cabs trips data in New York City. Niu~\etal~\cite{niu2019predicting} analyzed the ride request data provided by Didi Chuxing and proposed a region-partitioning-assisted LSTM order prediction model to help online ride-hailing platforms to better manage vehicular resources.

Lee~\etal~\cite{lee2008analysis} analyzed the passenger pickup patterns of taxi services in Jeju area on the location history data collected from taxi telematics system. They applied $k$-means clustering to obtain clusters based on the spatio-temporal pickup frequency, and used these cluster locations to recommend empty taxis the next pick-up location. Veleso~\etal~\cite{veloso2011urban} analyzed the taxi traces to identify relevant pickup and drop-off points and established relationships between them. They also characterized the scenario between the latest drop-off and next pickup and explored the possibility to predict the next pickup area type given the drop-off features. Authors in~\cite{hsueh2014effective,hwang2015effective} analyzed passengers' pickup- and drop-off locations to predict the most likely next pickup location given the current passenger drop-off locations and proposed a taxi recommender system.

Liu~\etal~\cite{liu2010uncovering} studied the cabdrivers' behaviour through their digital traces, operational patterns, and route choice behaviour for top and ordinary drivers, categorized based on their income.   Hu~\etal~\cite{hu2014exploring,hu2018taxi} explored the taxi driver operation behaviour by the measurements of activity space (pickup and drop-off locations) and the connection between different activity spaces for the different time duration. They also studied passengers' demand on a spatial-temporal distribution.  Jiang~\etal~\cite{jiang2017multi} conducted a multi-period analysis of taxi drivers' behaviours to extract passenger delivery and passenger searching trip information and evaluated taxi-drivers' working conditions. Su~\etal~\cite{su2018understanding} proposed a methodological framework to derive three indexes to measure pickup and drop-off dynamics from the taxicabs data in the city of Shenzhen.

Existing studies mainly focused on analyzing passengers' locations such as pickup and drop-off points, for different contexts such as ride-sharing, driver-passenger matching, next pick-up points prediction etc. However, to our knowledge, there is no work studying drivers' performance for passenger pick-ups at different \textit{drivers' locations} at the time of booking request. In this work, we analyze drivers' and passengers' locations at the time of pickup request for drivers' pickup performance based on the timely or late pickup.

\section{Preliminary Methods}\label{sec:preliminary}
The co-clustering algorithm, we implemented in this work, is based on the \textit{visual assessment of clustering tendency} (VAT)~\cite{bezdek2002vat} algorithm and some of its relative methods. Therefore, we briefly discuss them below.

\subsection{VAT and iVAT}
Consider a set of $N$ objects $O=\{o_{1},o_{2},...,o_{N} \}$ that can represent virtually anything such as web pages, documents, movies, drivers, locations, etc. Each object is represented by a $p$-dimensional feature vector, $\textbf{x}_{i} \in \mathbb{R}^{p}$ in a set of $X = \{\textbf{x}_{1},..,\textbf{x}_{N}\} \in \mathbb{R}^{p}$. Another way to represent the objects in $O$ is with 
a square dissimilarity matrix $D = [d_{ij}]$, where $d_{ij}$  represents dissimilarity between $o_{i}$ and $o_{j}$, computed using a chosen 'distance metric'.

The VAT algorithm~\cite{bezdek2002vat} reorders the dissimilarity matrix $D$ to $D^{*}$ using a modified Prim's algorithm that finds the \textit{minimum spanning tree} (MST) of a weighted undirected graph. Each pixel of the VAT image $I(D^{*})$, also called \textit{reordered dissimilarity image} (RDI), reflects the dissimilarity value between corresponding row and column objects. In a grayscale RDI image, $I(D^{*})$, white pixels represent high dissimilarity, while black represents low dissimilarity. Each object is exactly similar to itself, which results in zero-valued (black) diagonal elements, and non-zero valued off-diagonal elements in $I(D^{*})$. A dark block along the diagonal of RDI is a sub-matrix of "similar" dissimilarity values; therefore, when dark blocks appear along the diagonal of the RDI $I(D^{*})$,  they potentially represent different (ideally, $k$) clusters of objects that are relatively similar to each other.

\begin{figure}
\captionsetup[subfigure]{justification=centering}
\centering
\subfloat[Synthetic data  $N=5000$]{\includegraphics[width=0.12\textwidth]{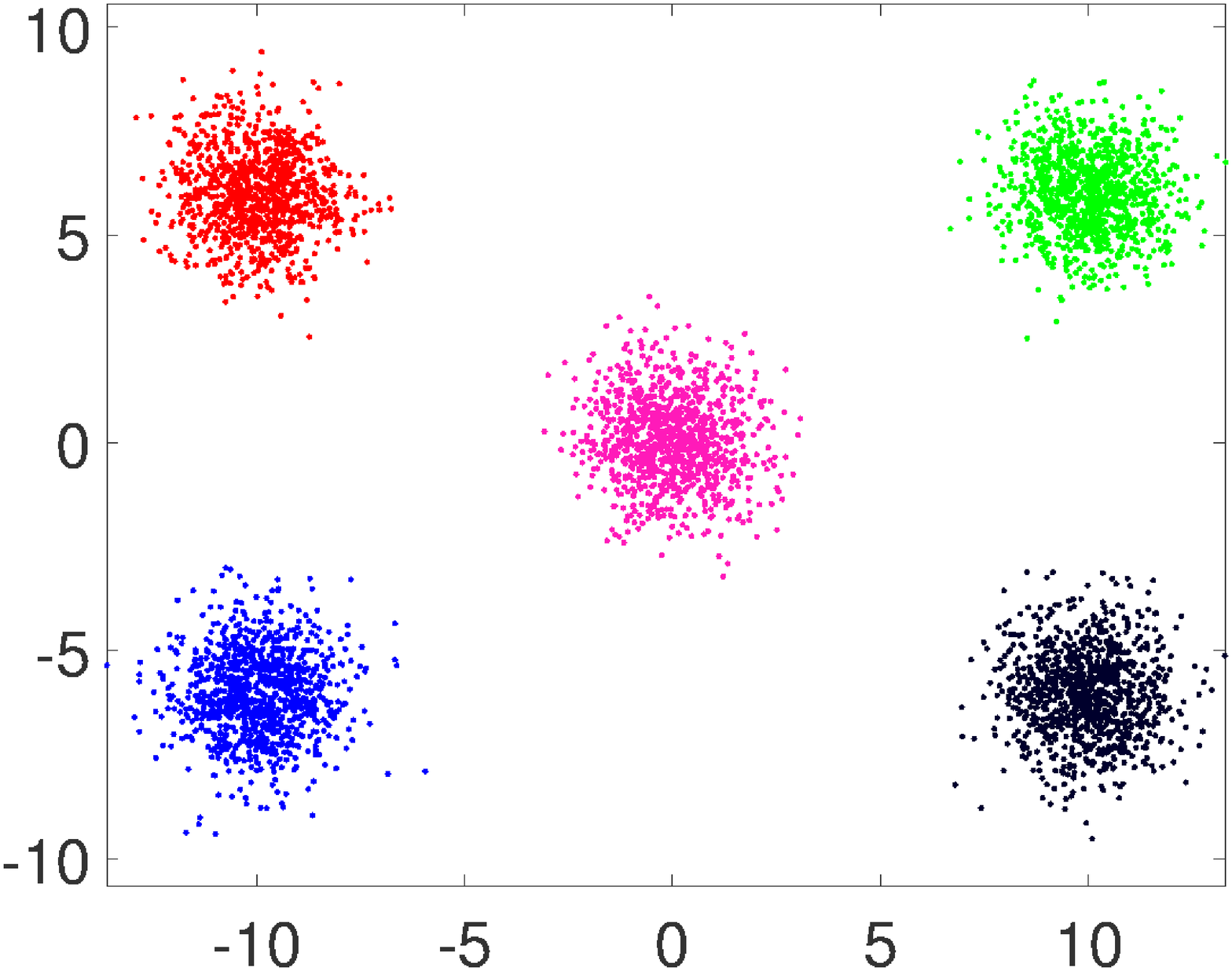}} \hfill
\subfloat[VAT for $N=5000$]{\includegraphics[width=0.12\textwidth]{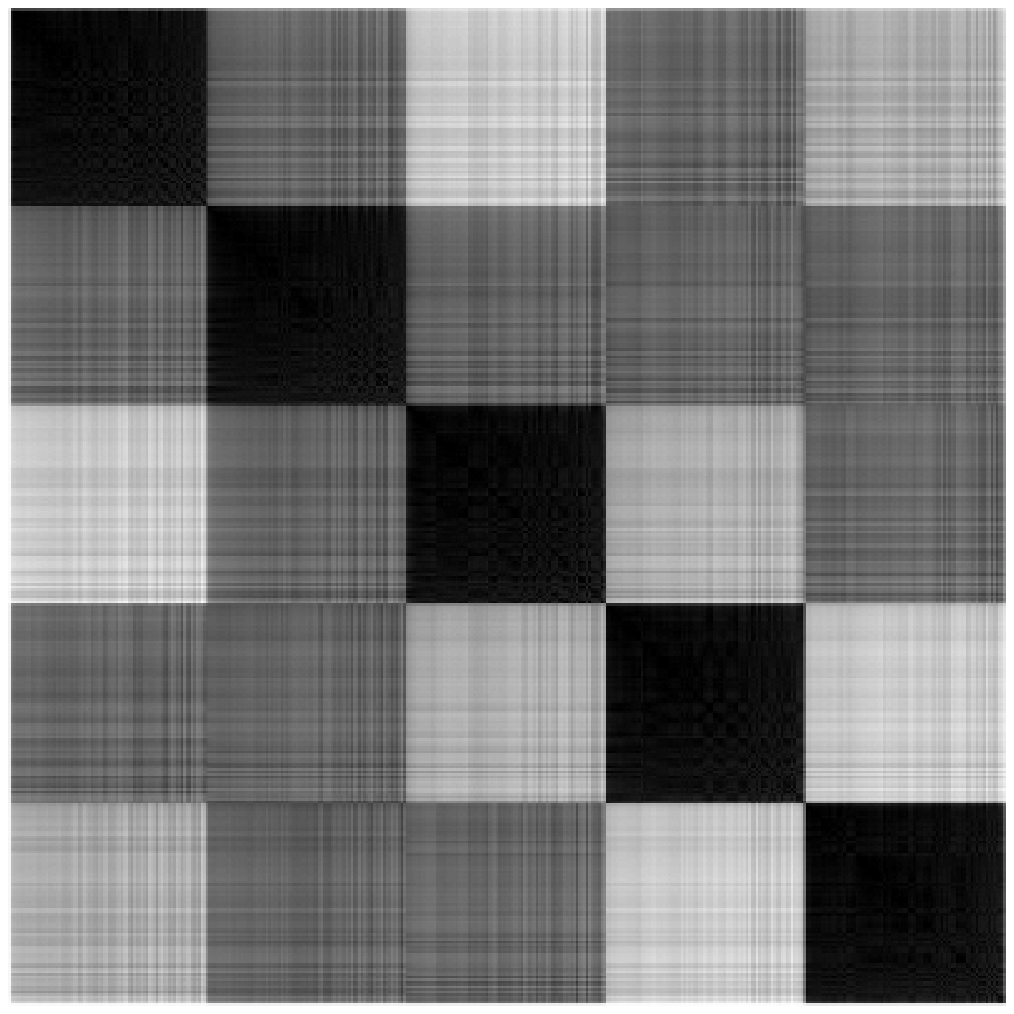}} \hfill
\subfloat[iVAT for $N=5000$]{\includegraphics[width=0.12\textwidth]{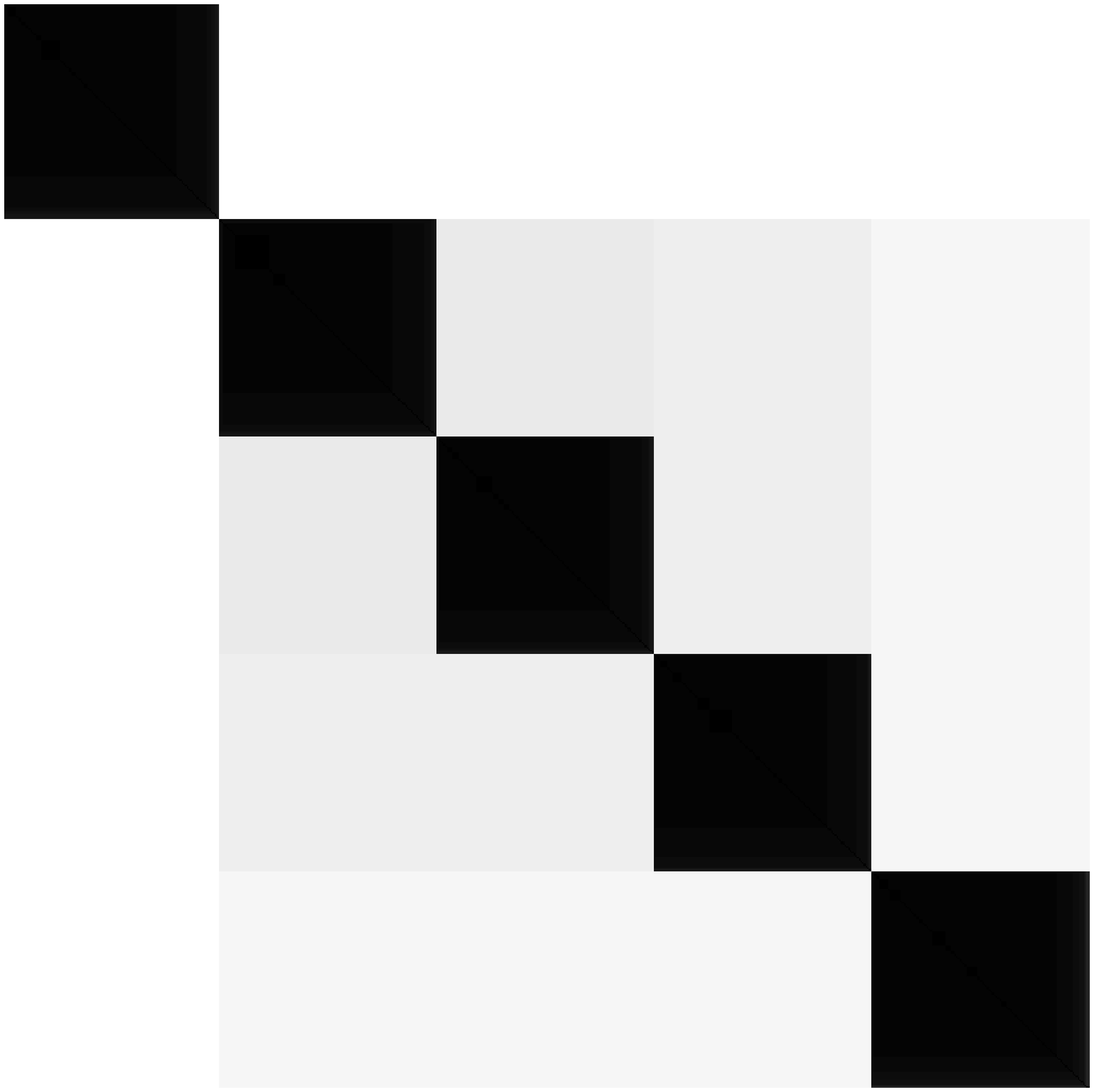}} \hfill
\subfloat[siVAT for $n=500$]{\includegraphics[width=0.12\textwidth]{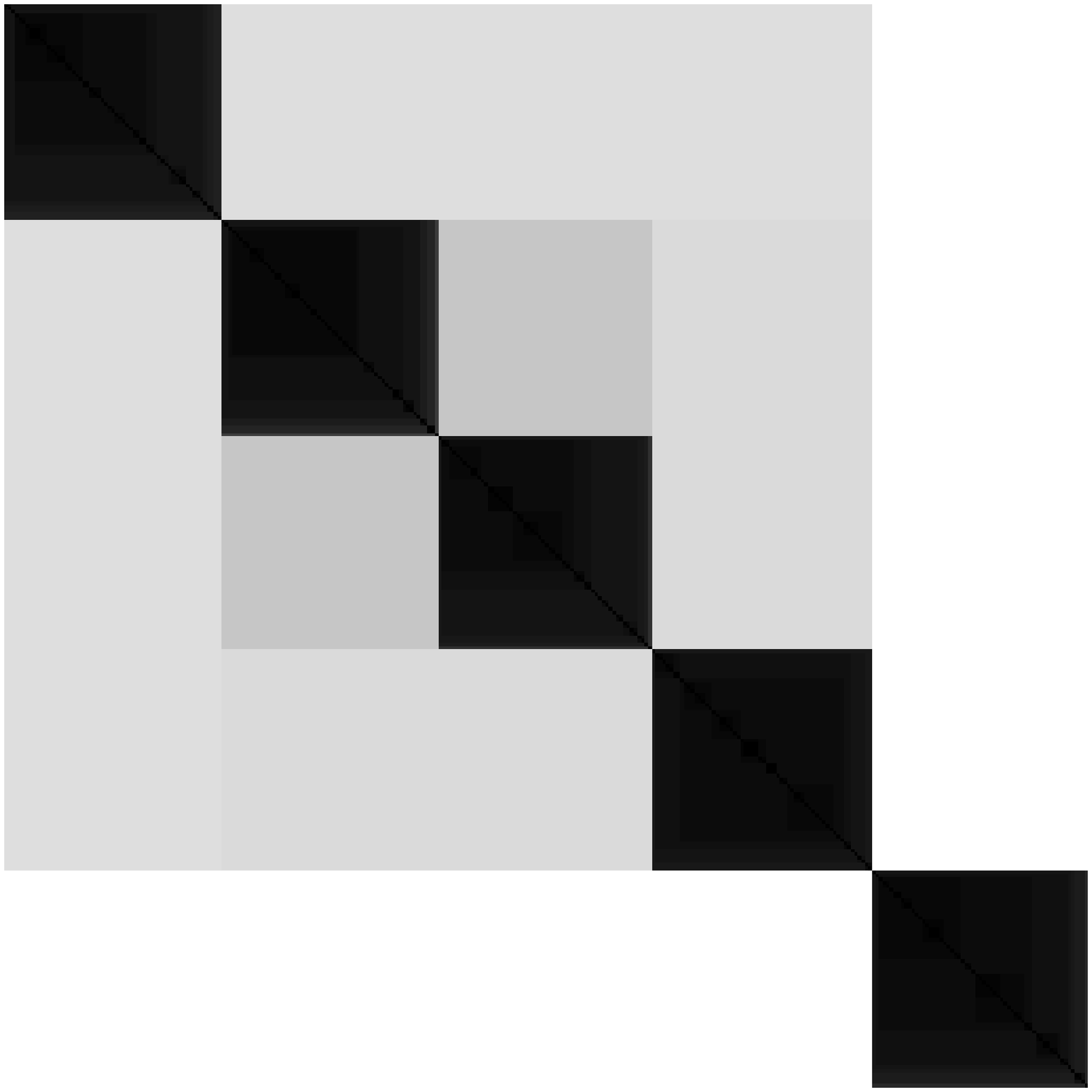}}
\caption{Data scatterplot, VAT, iVAT, and siVAT images for a 2D synthetic dataset.}
\label{Fig:VATiVATsiVATImage}
\end{figure}

Haven~\etal~\cite{havens2012efficient} proposed an improved VAT (iVAT) algorithm by replacing input distances $[d_{ij}]$ in  $D$ by path-based minimax distances $D^{'}= [d_{ij}^{'}]$, which are calculated as follows:
\begin{eqnarray}
d_{ij}^{'} = \operatorname*{min}_{r \in P_{ij}} \operatorname*{max}_{1<h<|r|} D_{N_{r[h]r[h+1]}},
\end{eqnarray}
where $r \in P_{ij}$ is an acyclic path in the set of all acyclic paths between objects $o_{i}$ and  $o_{j}$ (vertices $i$ and $j$) in $O$. An iVAT image (RDI) is represented by $I({D'}^{*})$. Essentially, iVAT is a distance transform that improves the visual contrast (sharpness) of the dark blocks along VAT image diagonal.

VAT and iVAT can handle only upto moderately sized data sets (with a few tens of thousands of data points) due to its $O(N^2)$ computational complexity. For big data, Hathaway~\etal~\cite{hathaway2006scalable} developed a scalable version of VAT/iVAT called \emph{scalable VAT/iVAT} (sVAT/siVAT), which first extracts a sample of (approximately) size $n$ ($n<<N$) from the big data $X$ using a smart sampling scheme~\cite{johnson1990minimax}, and then applies VAT to the (small) distance matrix computed from the extracted sample. The sample is chosen so that it (hopefully) contains a cluster structure similar to the full dataset. This is obtained by first picking a set of $k'$ distinguished (furthest from each other) objects using maximin sampling~\cite{johnson1990minimax}, selected to provide a representation of each cluster. Then, the remainder of the sample is built by choosing additional data near each of the distinguished objects using random sampling. This sampling scheme is called maximin random sampling (MMRS)~\cite{johnson1990minimax}. 

\Fig~\ref{Fig:VATiVATsiVATImage} illustrates VAT, iVAT, and siVAT for a 2D synthetic dataset ($N=5000$). While both VAT  and iVAT images show five dark blocks along the diagonal corresponding to the five clusters in the dataset, dark blocks in the iVAT image are much clearer than the VAT image. Besides, siVAT allows us to create a similar image by sampling (using MMRS) only $n=500$ points (0.05~\% of the total datapoints).

\subsection{co-VAT and sco-VAT}
VAT and iVAT algorithms can only handle square dissimilarity matrix (or, in more general terms, relational) matrix i.e. when the row and column objects in $D$ comprise the same type of elements of $O$. A more general form of relational data is rectangular matrix between $M$ row objects, $O_r$, and $N$ column objects, $O_c$. An example is a word-document analysis where rows correspond to $M$ documents, the columns correspond to $N$ words, and matrix entries correspond to occurrence measures of words in documents. Another example is gene expression data where rows represent genes, and columns represent tissue samples or conditions. In rectangular relation data, there can be groups of similar objects that are composed of only row objects, only column objects, and only mixed objects that are often called co-clusters~\cite{dhillon2003information}. 

\begin{algorithm}
\caption{sco-iVAT}\label{Algo1}
\scalebox{0.8}{
\begin{minipage}{0.6\textwidth}
\begin{algorithmic}
\Statex \textbf{Input:}  $D_{M \times N}$ - $M \times N$ relational data matrix, $M$ and $N$ are large; $m$- number of samples from rows objects; and $n$- number of samples in column objects; k'- number of distinguished objects. 
\Statex \textbf{Output:} Row iVAT image $I({[{D'_r}^{*}]}_{m \times m})$; column iVAT image $I({[{D'_c}^{*}]}_{n \times n})$; and RRI $I({D}_{m \times n}^{*})$;
\end{algorithmic}
\begin{algorithmic}[1]
\State Consider the rows and columns of $D$ as the feature vectors representing $M$ row objects and $N$ column objects, respectively.
\State Apply MMRS sampling to the set of $M$ row objects ($M$~ $N$-dimensional feature vectors) returning $m$ sampled row objects, and build ${[D_r]}_{m \times m}$.
\State Apply MMRS sampling to the set of $N$ column objects ($N$~  $M$-dimensional feature vectors)  returning $n$ sampled column objects, and build ${[D_c]}_{n \times n}$.
\State Build  ${D}_{m \times n}$ by extracting $m$ sampled rows and $n$ sampled columns from $D_{M \times N}$.
\State Apply iVAT to ${[D_r]}_{m \times m}$ to generate permutation array for row objects, $RP= \{P_1,P_2,...,P_m \}$ and obtain row-RDI $I({[{D'_r}^{*}]}_{m \times m})$

\State Apply iVAT to ${[D_c]}_{n \times n}$ to generate permutation array for row objects, $CP= \{C_1,C_2,...,C_n \}$ 
and obtain column-RDI $I({[{D'_c}^{*}]}_{n \times n})$.

\State Reorder rows and columns of ${D}_{m \times n}$  based on permutation array $RP$ and $CP$, respectively, to obtain  reordered relational matrix ${D}_{m \times n}^{*}$ and image $I({D}_{m \times n}^{*})$. 
\Statex // \textit{Following steps are optional} //
\State  Obtain aligned clusters (say $k$) from sco-iVAT image  $I({D'}_{m \times n}^{*})$.
\State Label remaining $(M-m)$ row and $(N-n)$ column objects by giving them the label of their nearby sampled row and column objects, respectively.
\State Obtain co-cluster by selecting the group of column objects, for each group of row objects.
\end{algorithmic}
\end{minipage}%
}
\end{algorithm}

To tackle rectangular relational data, Havens~\etal~\cite{havens2010new} proposed an approach for visually assessing cluster tendency for the objects represented by rectangular relational data matrix $D$. The coVAT technique proposed in~\cite{havens2012new} generates a reordering of the rows and column indices of $D$ by applying VAT algorithm to dissimilarity matrices $D_r$ and $D_c$, respectively. The dissimilarity matrices $D_r$ and $D_c$ are computed using feature vectors of objects $O_r$ and $O_c$, respectively. Based on this row and column reordering, the rows and columns of rectangular data $D$ are reordered to obtain a reordered relational matrix $D^{*}$ and its co-VAT image $I(D^{*})$. Just as with VAT, dark blocks in $I(D^{*})$ (not along any diagonal, and not necessarily square) suggest the existence of co-clusters.

Similar to VAT and iVAT, the co-VAT algorithm also suffers
from high memory requirements and computational complexity
as the data size increases. To address this issue, Park~\etal~\cite{park2009visualization} utilized the sVAT sampling across both row and column objects to extend co-VAT to handle large rectangular data and named their algorithm scalable-coVAT (sco-VAT).

\section{sco-iVAT}\label{sec:sco-iVAT}
In this article, we developed an improved version of scalable coVAT, which we call sco-iVAT, and apply it on a driver-location-performance relational data to obtain a representative sample of drivers and locations, and subsequently, to obtain possible clusters and co-clusters from their sco-iVAT image. The pseudocode of sco-iVAT is given in Algorithm 1. In the first step, rows and columns of input relational matrix, $D$, are interpreted as $M$ row objects of $N$-dimensional feature vectors and $N$ column objects of $M$-dimensional feature vectors, respectively.  Then, MMRS sampling is applied to these $M$ row objects and $N$ column objects to obtain $m$ sampled objects from $M$ rows and $n$ sampled objects from $N$ columns (lines 2-4). Subsequently, iVAT is applied to ${[D_r]}_{m \times m}$ and ${[D_c]}_{n \times n}$ (computed from $m$ sampled row and $n$ column objects)  to obtain reordering of sampled rows and column objects (lines 5-6) which, in turn, yield us reordered matrix ${D'}_{m \times n}^{*}$ (line 7). 

The sco-iVAT provides the visual estimate\footnote{There exists several methods in the literature~\cite{wang2009automatically} to automatically determine the number of clusters from iVAT images without human intervention.} of possible clusters and co-clusters from row iVAT image $I({[{D'_r}^{*}]}_{m \times m})$; column iVAT image $I({[{D'_c}^{*}]}_{n \times n})$; and \textit{reordered relational matrix image} (RRI)  $I({D}_{m \times n}^{*})$, respectively (line 8). Since single-linkage clusters are always diagonally aligned in VAT/iVAT ordered images~\cite{havens2013scalable}, so (say $k$) aligned clusters can be obtained by cutting the largest $(k-1)$ edges in the MST for row and column objects (line 9). Once the sampled row and column objects are labelled, the remaining $(M-m)$ row objects and $(N-n)$ column objects can be labelled by extending the label of the sampled row and column objects to them using nearest prototyping rule. Subsequently, co-clusters points can be obtained by selecting the datapoints in the same group across columns, for each group of rows objects (line 10).

\begin{figure}
\captionsetup[subfigure]{justification=centering}
\centering
\subfloat[Dataset $X_1$, $M=4000, N=3000$ ]{\includegraphics[width=0.26\textwidth]{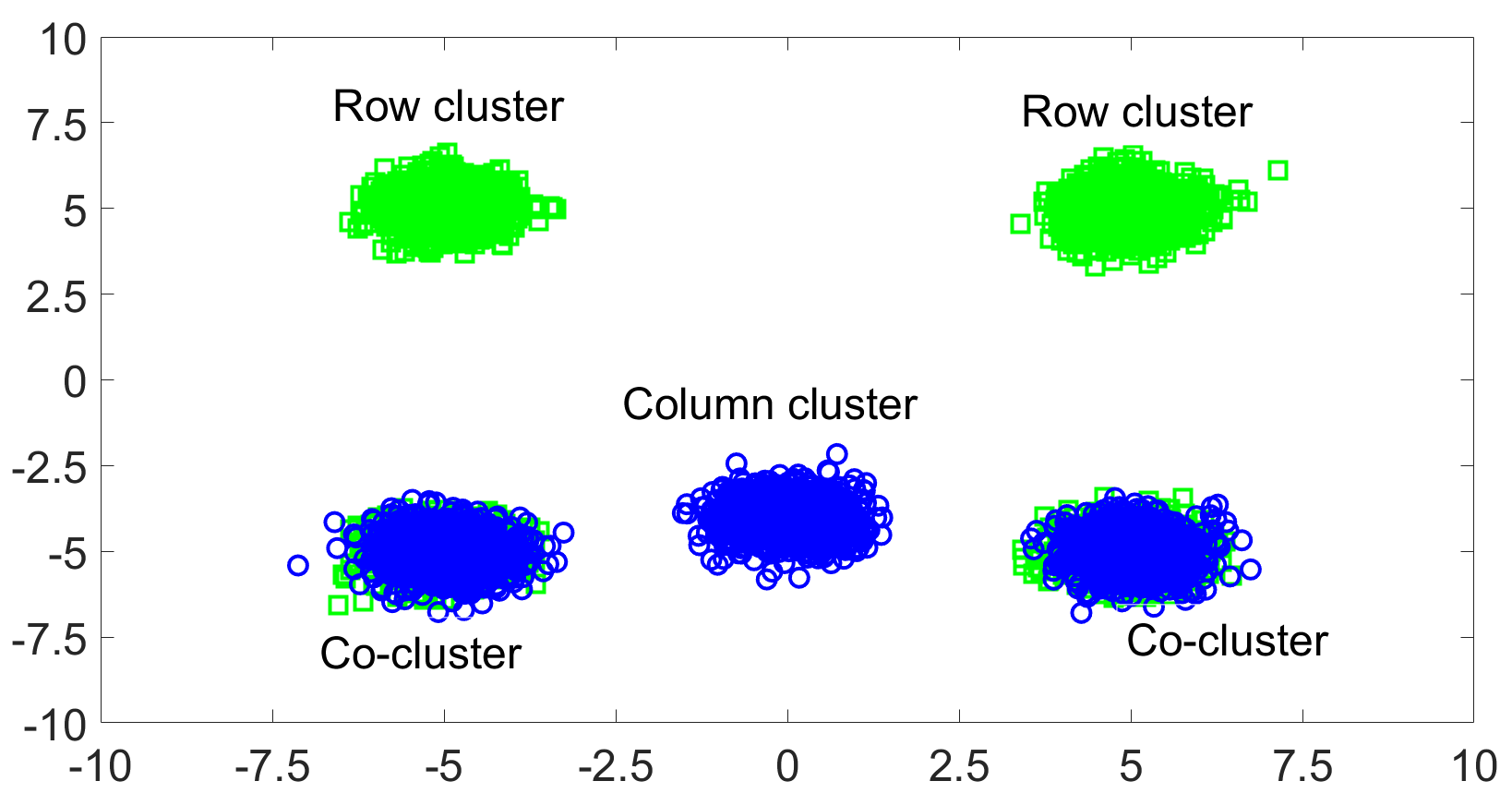}} 
\subfloat[$I({[{D'_r}^{*}]}_{105 \times 105})$]{\includegraphics[width=0.12\textwidth]{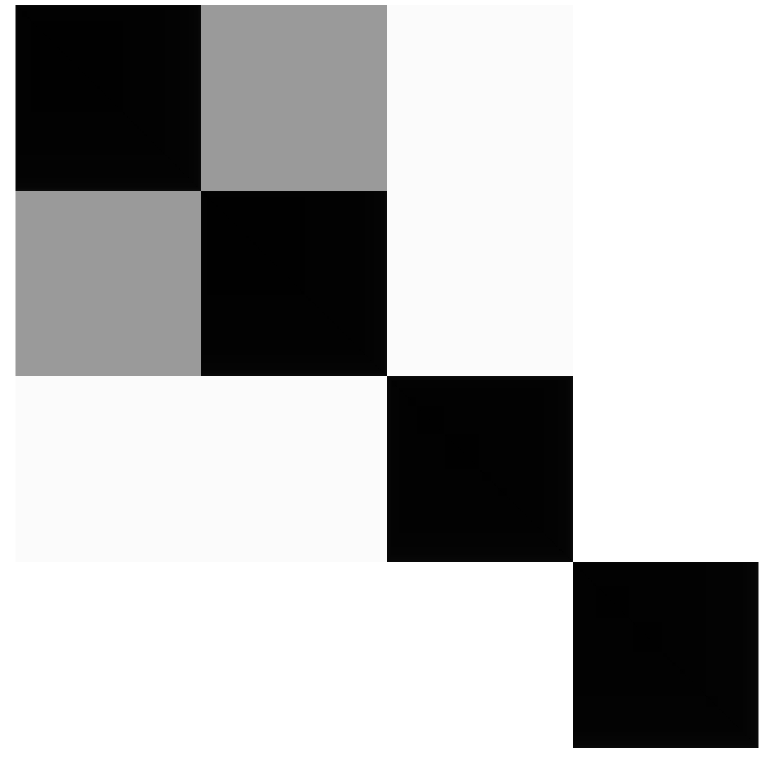}} 
\subfloat[$I({[{D'_c}^{*}]}_{36 \times 36})$]{\includegraphics[width=0.12\textwidth]{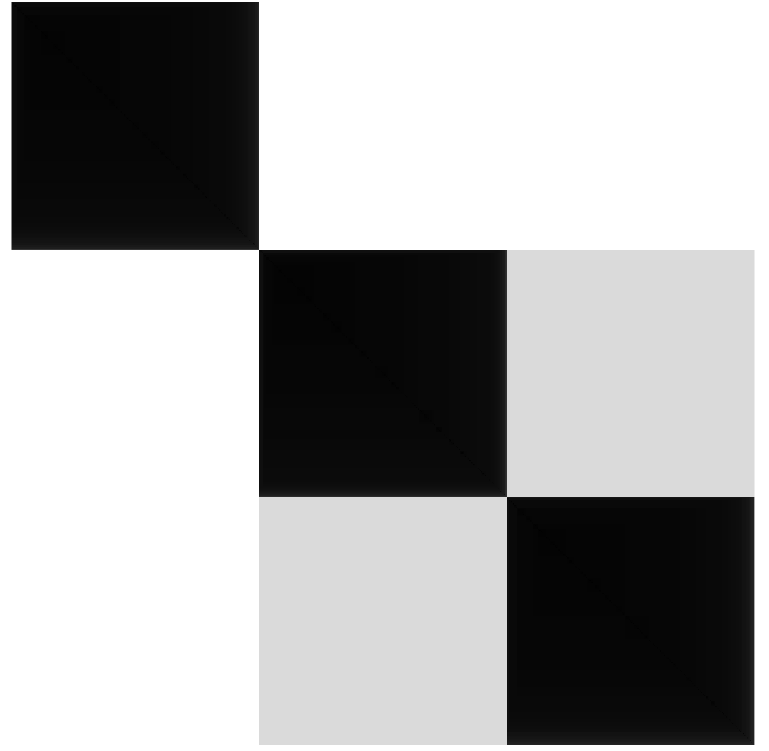}}\\
\subfloat[Dissimilarity matrix, I($D_{M \times N}$)]{\includegraphics[width=0.25\textwidth]{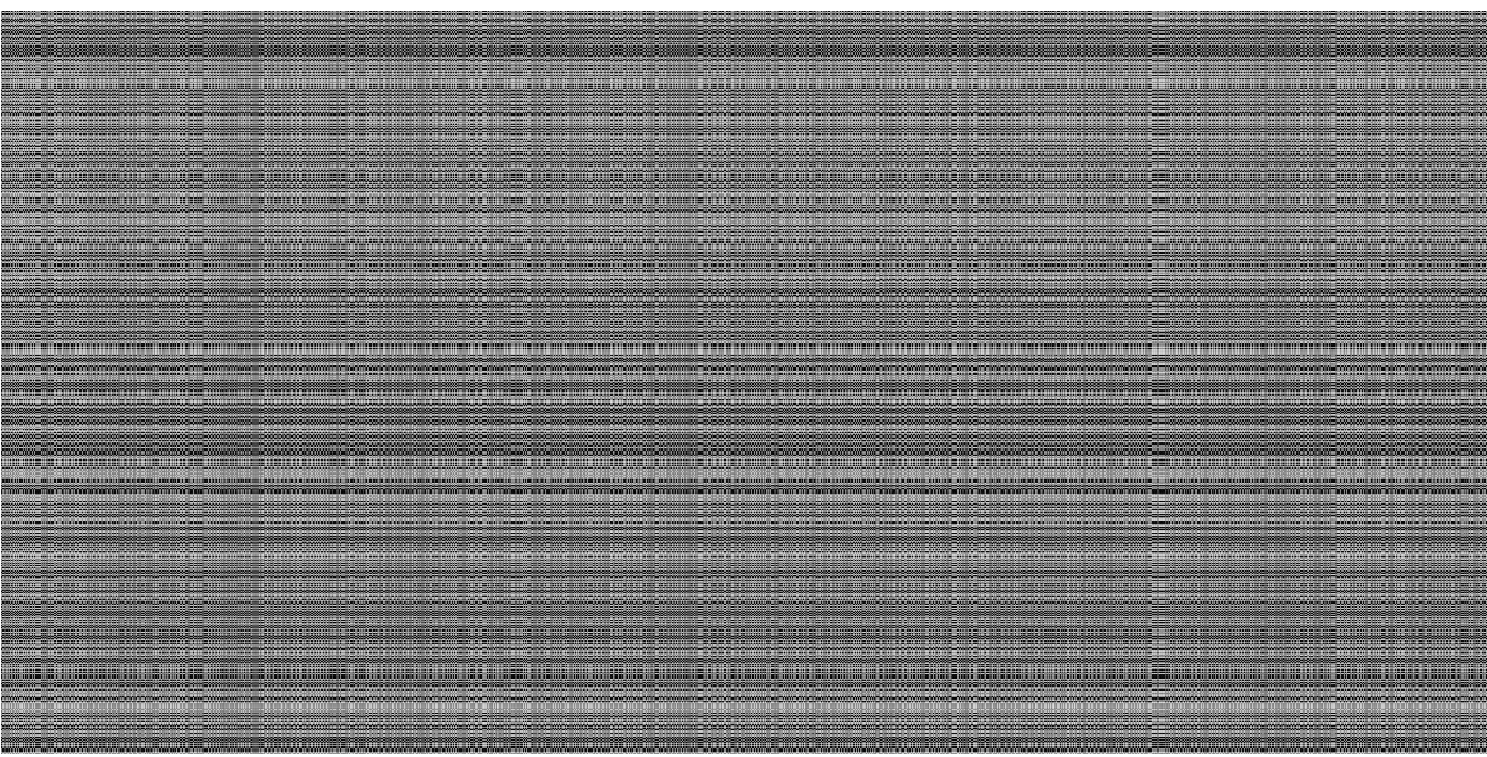}} 
\subfloat[$I({D}_{105 \times 36}^{*})$]{\includegraphics[width=0.25\textwidth]{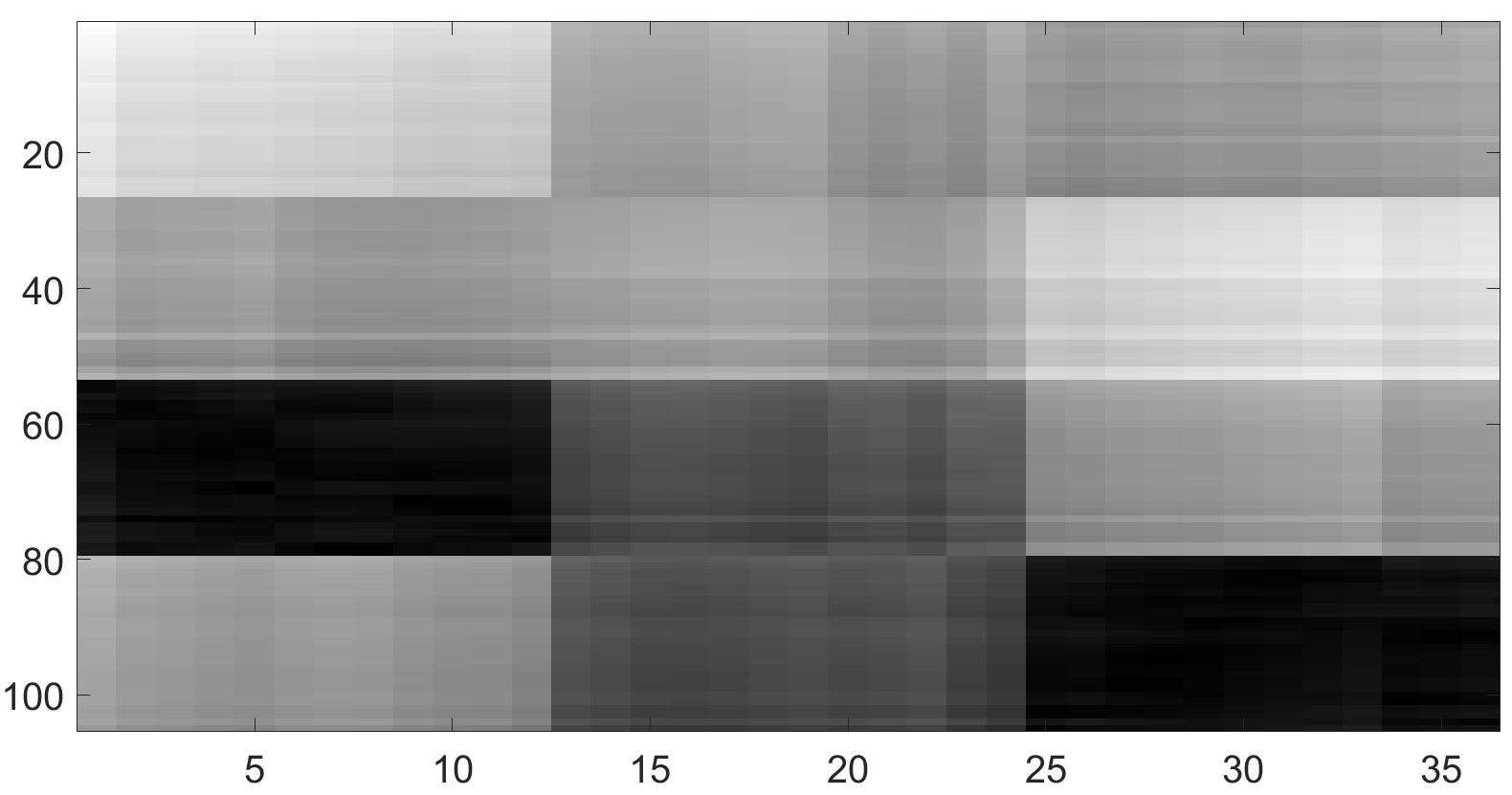}}
\caption{Example 1: An example illustrating sco-iVAT with a synthetic data with mixed row and column objects.}
\label{Fig:scoiVAT_Example1}
\end{figure}

\begin{figure}
\captionsetup[subfigure]{justification=centering}
\centering
\subfloat[Relational Matrix, $M=10000, N=8000$ ]{\includegraphics[width=0.25\textwidth]{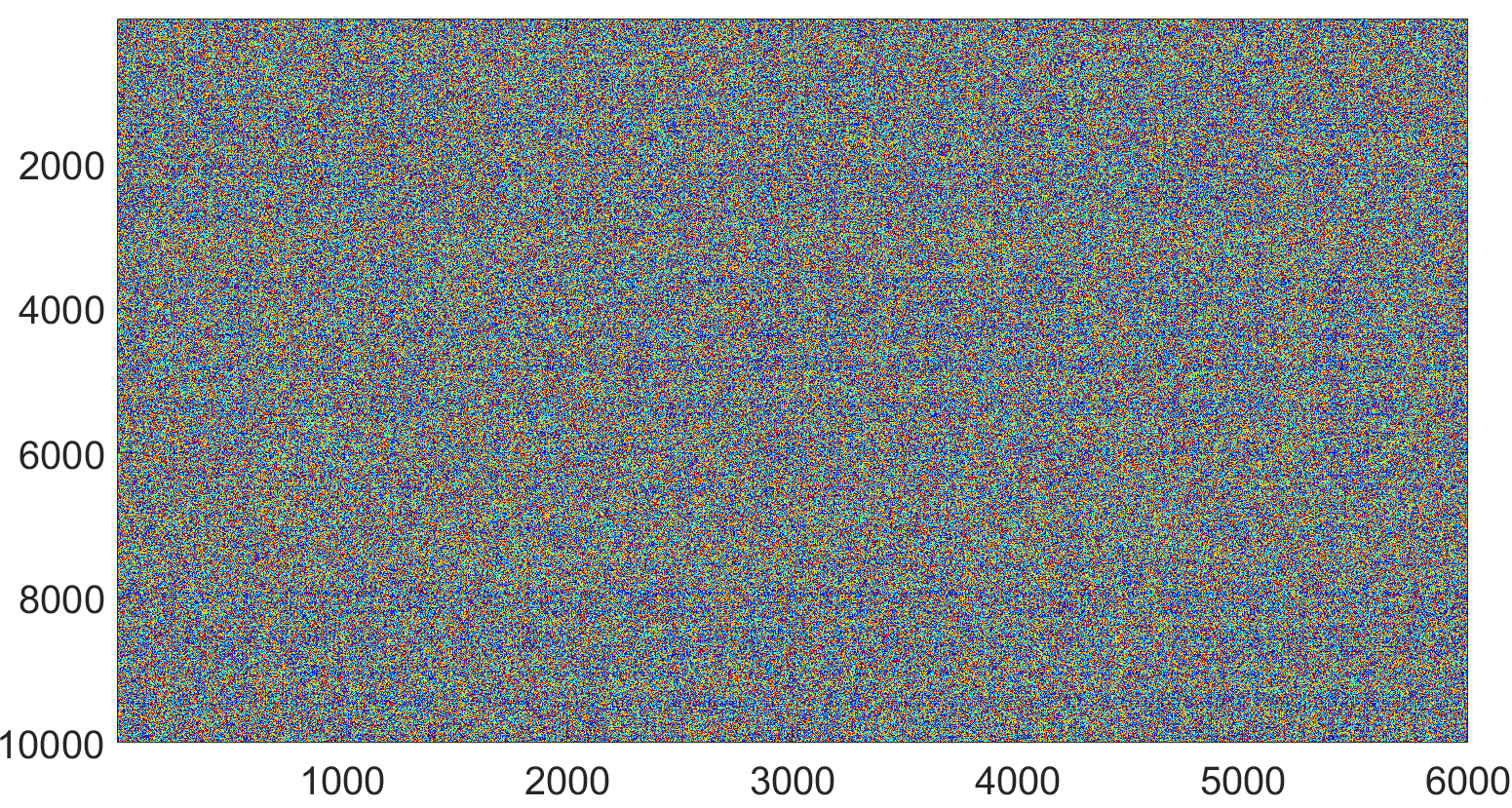}} 
\subfloat[sco-iVAT reordered relational matrix, I($D_{105 \times 84}$)]{\includegraphics[width=0.25\textwidth]{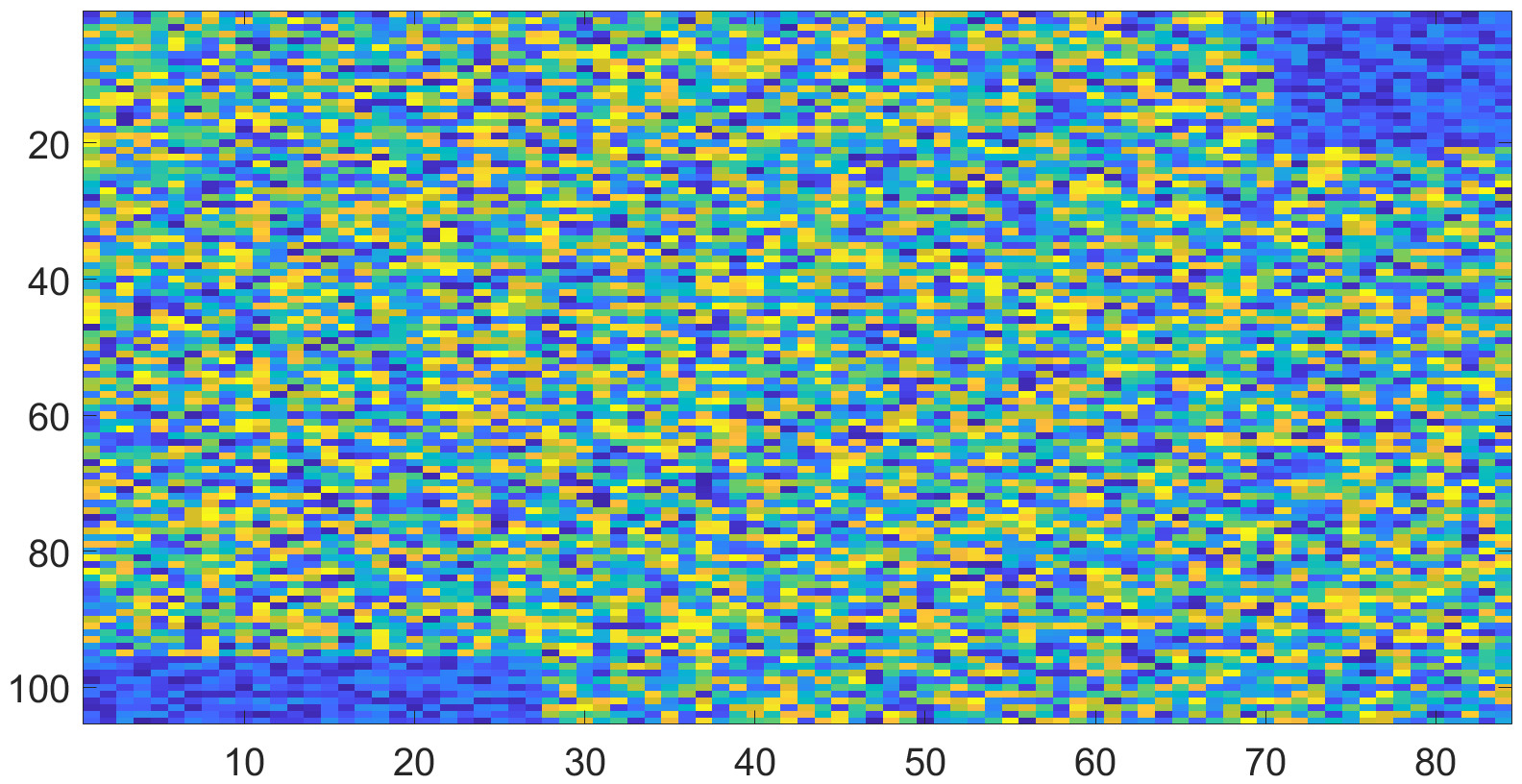}}
\caption{Example 2: An example illustrating sco-iVAT with a relational data matrix having two co-clusters.}
\label{Fig:scoiVAT_Example2}
\end{figure}

\textbf{Example 1:}
\Fig~\ref{Fig:scoiVAT_Example1} illustrates the sco-iVAT for a synthetic dataset $X_1$ having $M= 4000$ row objects and $N=3000$ column objects, where each row and column cluster is normally distributed with its own mean and covariance matrix.
So, $X_1$ has $4$ pure row-clusters (in green), $3$ pure column-clusters (in blue), and $2$ co-clusters (mixed). The dark blocks in row and column RDIs, obtained from sco-iVAT, confirms $4$ and $3$ pure clusters, respectively, as seen in \Figs~\ref{Fig:scoiVAT_Example1} (b)(c). While the grayscale image (view (d)) of randomly permuted rectangular dissimilarity matrix (constructed from rows and column objects) indicates no clusters in the data,  the dark blocks in its reordered dissimilarity matrix image ((view (e)) for sampled row ($m=105$) and column (n=$36$) objects clearly suggests $2$ co-clusters.\\
\textbf{Example 2:}
\Fig~\ref{Fig:scoiVAT_Example2} illustrates the sco-iVAT for a
relational data matrix $D_{10000 \times 8000}$, constructed using uniformly distributed random numbers in interval $[0  ~3]$. Two small matrices of sizes, $1000 \times 2000$  and $2000 \times 1000$, respectively, were generated using a different uniform distribution in interval $[0  ~1]$ and inserted in the original relational matrix at random rows and columns. While the color image of relational matrix does not indicate any clusters in~\Fig~\ref{Fig:scoiVAT_Example2} (a), the two rectangular blocks (in blue) in reordered relational matrix image  (\Fig~\ref{Fig:scoiVAT_Example2} (b)) of the sampled rows ($m$=105) and columns ($n$=84)  clearly indicates $2$ notable co-clusters, corresponding to the two different distributions across certain rows and columns.

\section{Methodology}\label{sec:methodology}
In this section, first, we will briefly describe the Grab booking dataset~\cite{huang2019grab} that we used in this work. Due to Grab's business interests, we reduce some level of detail to present only aggregated or secondary measures. Then, we discuss the pre-processing steps to prepare data for our analysis. Next, we discuss the feature extraction from the pre-processed booking data that we feed into co-clustering (sco-iVAT) and classification model. Finally, we present the results and driver scoring mechanisms for booking assignments. The results presented here must not be assumed to indicate any of the company's business interests.

\subsection{Dataset}
The dataset is sampled from Grab's booking records and their drivers' trajectories in Singapore with personal information encrypted and real start and end locations removed. This sampled dataset (after pre-processing) contains $127732$ booking records of $48205$ drivers over one month, with each booking having the following features:
\begin{itemize}
    \item booking ID (configured with hashes)
    \item driver ID (configured with hashes)
    \item booking acceptance timestamp
    \item \textit{driver's location} at the time of booking acceptance
    \item assigned driver's GPS pings with timestamps  
    \item passenger pickup location 
    \item passenger pickup timestamp
    \item ETA for passenger pickup
    \item ATA for passenger pickup.
\end{itemize}

The GPS pings were collected from drivers' smartphones during driving. Each GPS ping of a driver is associated with  booking ID, latitude, longitude, timestamp, accuracy level, and speed information. The GPS sampling rate is $5$ seconds. The accuracy level indicates the accuracy of GPS pings in the horizontal plane~\cite{huang2019grab}.  The accuracy level indicates the radius within which the location confidence is $68$\% i.e., given a circle centred at the reported latitude and longitude, and with a radius equal to the accuracy level, then there is a 68\% probability that the true location is inside the circle. Note that we will not be using entire GPS trajectories in our analysis.

\begin{figure}
\captionsetup[subfigure]{justification=centering}
\centering
\subfloat[day-wise distribution]{\includegraphics[width=0.48\columnwidth]{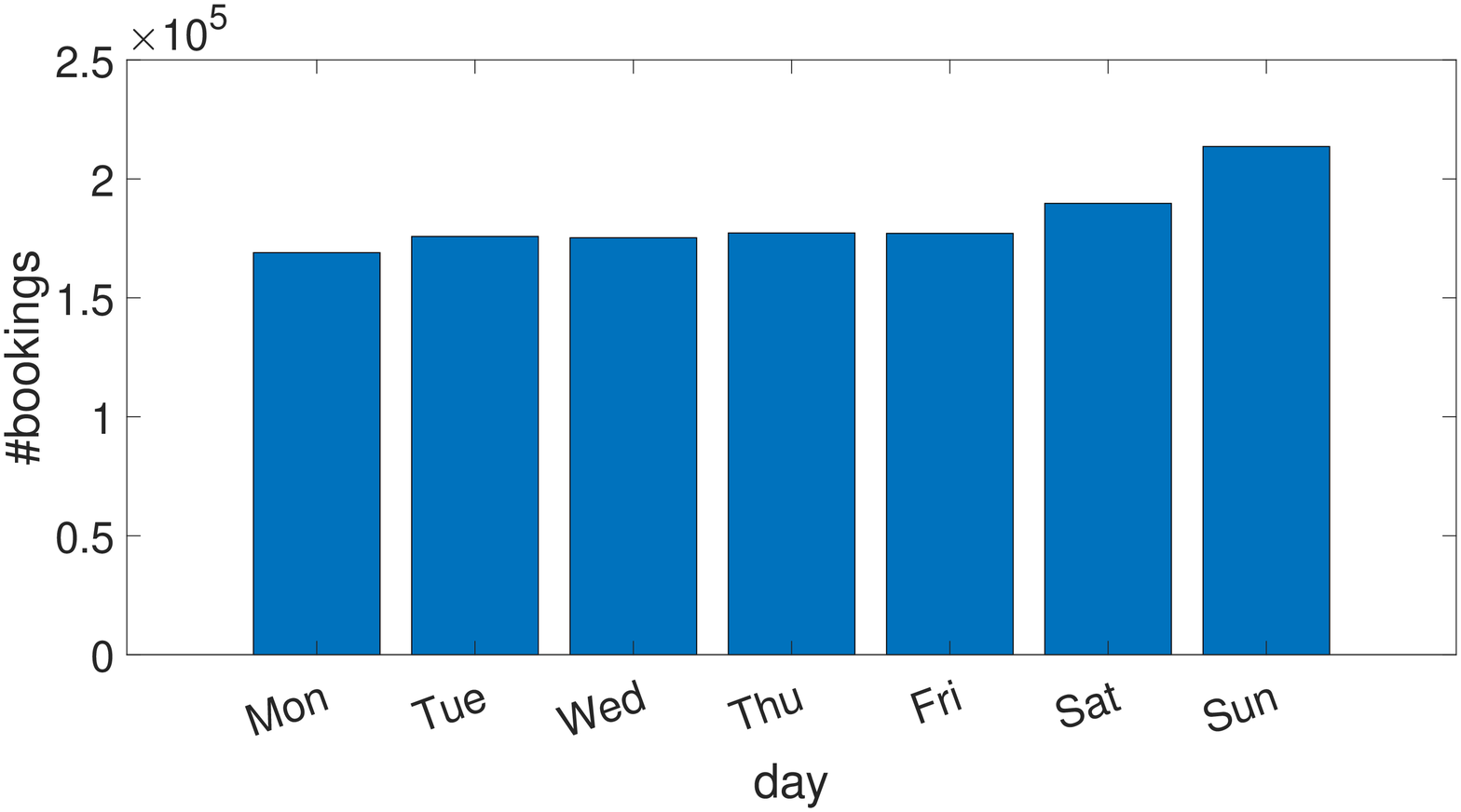}} 
\subfloat[hourgroup-wise distribution]{\includegraphics[width=0.48\columnwidth]{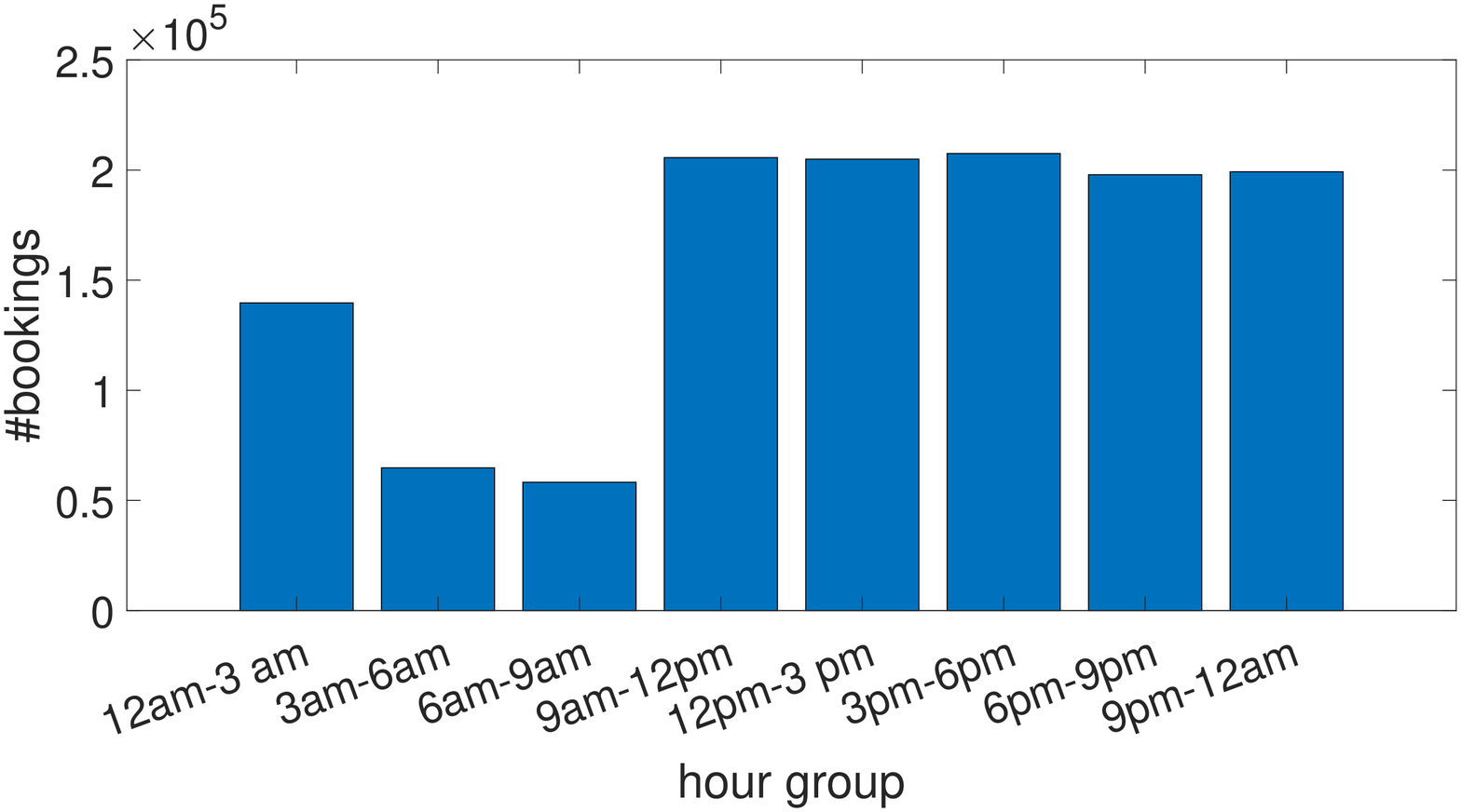}}
\caption{Day- and hourgroup-wise distribution of bookings}%
\label{Fig:BookingDistribution}
\end{figure}
 \setlength{\textfloatsep}{-1pt}

\subsection{Pre-processing}
The raw GPS coordinates were map-matched using HMM-based map matching~\cite{huang2019grab,newson2009hidden}  
 algorithm to infer the most likely road segment onto which each GPS point belongs. There were some booking records with noisy ETA due to inaccurate GPS points. These noisy ETA records were filtered as follows: first, we estimated a speed based on the travelled distance and ETA for each record, and then filtered the records with speed falling within $[0~110]$ km/hour range. No ride-sharing or incomplete bookings were considered in the data. 

It is desired to understand the driver pickup performance in small areas, however, aggregation of different features by smaller areas or GPS coordinates is not feasible. So, we need to choose an area size for which aggregation is feasible and sufficiently accurate for the drivers. In this work, we chose geohash, a hierarchical geocoding system based spatial indexing, that hierarchically divides a geographical area into grid-shaped buckets with arbitrary precision. The size of the grid is determined by the number of characters used in the geohash code. We divided the entire Singapore map into $845$  grids by converting GPS coordinates into six character precision geohashes ($1.2km \times 0.6km$) and aggregated the booking features for each geohash grid.  We call \textit{driver's location} geohash as "driverGh" and \textit{pickup location} geohash as "pickupGh". 

To include temporal information, we extract the day of the week (DoW) i.e., weekday or weekend, and time of the day, during which the booking was made. Specifically, we divide 24 hours into 8 hour groups of 3 hours each i.e., 12am-3am, 3am-6am, 6am-9am, 9am-12pm, 12pm-3pm, 3pm-6pm, 6pm-9pm, 9pm-12am.~\Fig~\ref{Fig:BookingDistribution} shows the day- and hourgroup-wise number of bookings distribution of the sampled data.

\subsection{Feature Extraction}
Different features contain various type of useful spatio-temporal information about drivers and their bookings. Initially, we decided to extract all possible features and then identify the most informative and predictive ones using a feature selection method. First, for each booking, we compute the following features that represents driver's pickup performance for a booking: 
\begin{enumerate}
\item diff\_eta\_ata: Difference between ATA and ETA i.e., $(ATA-ETA)$. The higher the difference, the poorer the performance of a driver in terms of pickup time.
\item is\_late\_pickup: A Boolean variable which is 1 if a driver is late for passenger pickup, else 0. In our work, a pickup is considered late if $(ATA-ETA) > 5 mins$, else it is considered a timely pickup.
\end{enumerate}
We compute two additional features from driver trajectories for each booking and use them in classification task. 
\begin{enumerate}[resume]
\item start\_ata: Actual time taken by a driver to travel the first $50$ meters after accepting the booking.
\item end\_ata: Actual time taken by a driver to travel the last $50$ meters during passenger pickup.
\end{enumerate}
The motivation of extracting these two features is discussed in Section~\ref{sec:timelypickup}. Then, we aggregate the  four features mentioned above over all bookings in our data to compute the  (i) total\_bookings, (ii)  avg\_diff\_ata\_eta (avg is the acronym for \textit{average}), (iii) late pickup rate (LPR) in (\%)  (iV) avg\_start\_ata, and (v) avg\_end\_ata for:
\begin{enumerate}
\item each driver ID
\item each driver ID \textbf{and} weekday/weekend
\item each driver ID \textbf{and} hourgroup
\item each driverGh 
\item each driverGh \textbf{and} weekday/weekend
\item each driverGh \textbf{and} hourgroup
\item each pickupGh
\item each pickupGh \textbf{and} weekday/weekend
\item each pickupGh \textbf{and} hourgroup
\item each driver ID \textbf{and} driverGh 
\item each driver ID \textbf{and} driverGh \textbf{and} weekday/weekend
\item each driver ID \textbf{and} driverGh \textbf{and} hourgroup
\item each driver ID \textbf{and} pickupGh 
\item each driver ID \textbf{and} pickupGh \textbf{and} weekday/weekend
\item each driver ID \textbf{and} pickupGh \textbf{and} hourgroup.
\end{enumerate}
We call these features as \textit{aggregated\_features\.} that are used in our subsequent analysis. 

\begin{figure*}[ht]%
\captionsetup[subfigure]{justification=centering}
\centering
\subfloat[Relational data matrix image $I(D_{48205 \times 845})$]{\includegraphics[width=0.8\textwidth]{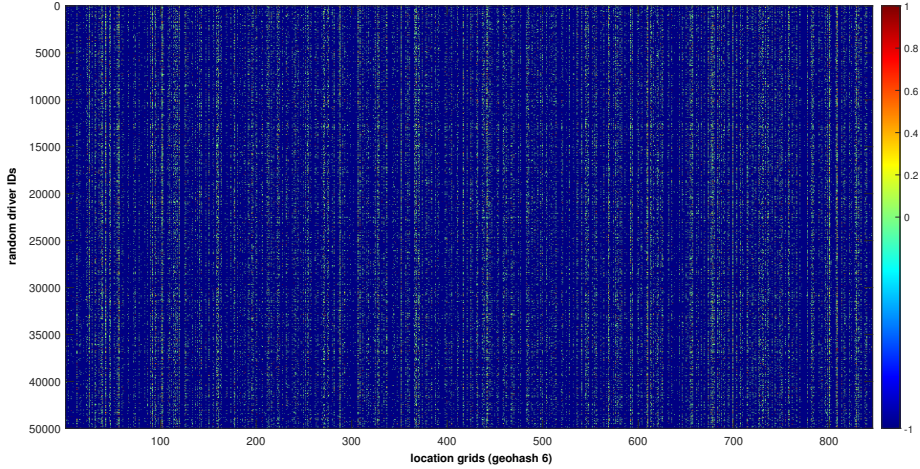}} \\
\subfloat[Reordered data matrix image $I({D}_{1069 \times 845}^{*})$ after co-clustering.]{\includegraphics[width=0.8\textwidth]{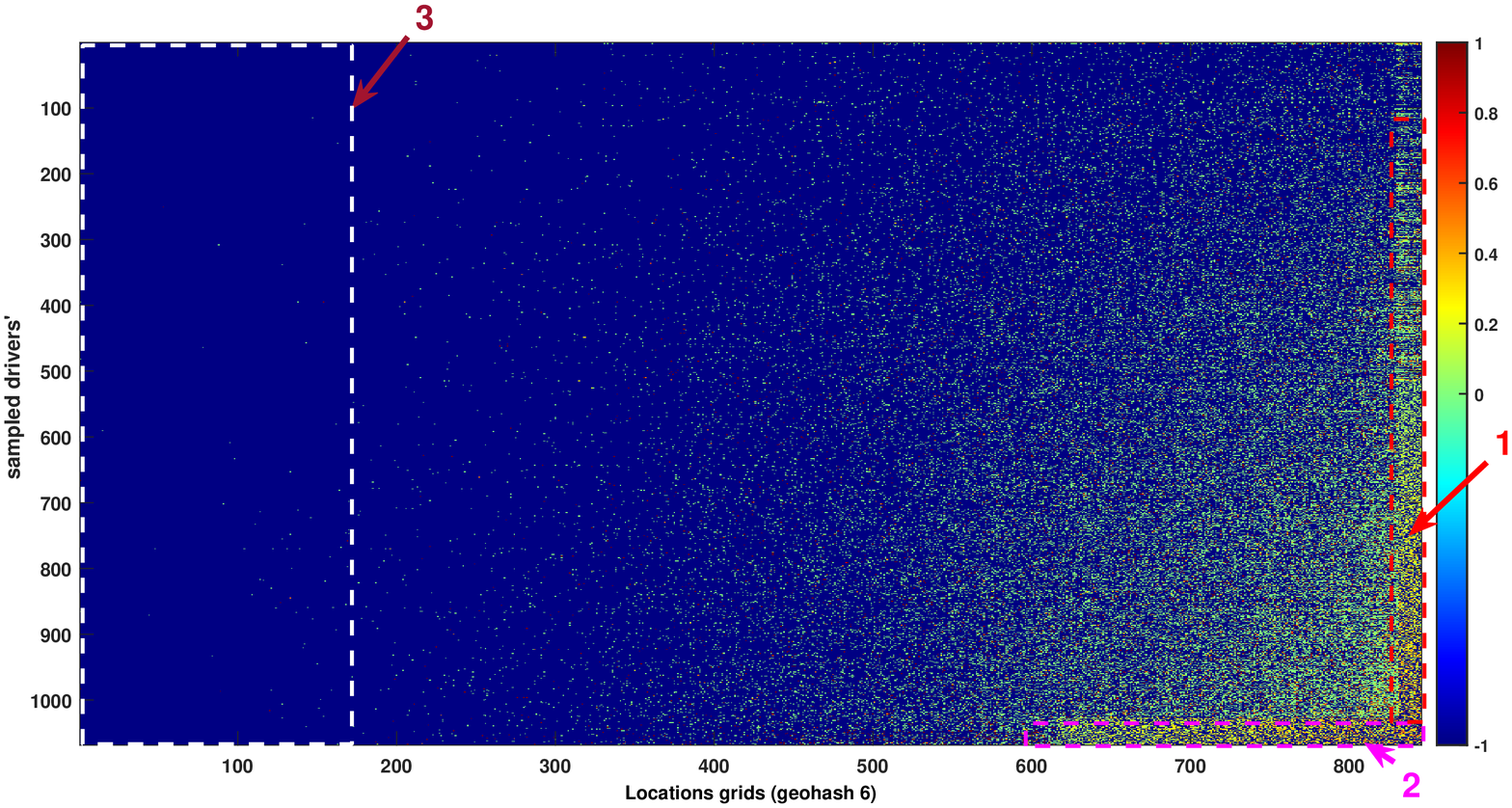}}
\caption{Driver-Location-Performance based relational matrix before and after applying sco-iVAT algorithm.}%
\label{Fig:CoclusteringResults}
\end{figure*}

\begin{figure*}
\captionsetup[subfigure]{justification=centering}
\centering
\subfloat[]{\includegraphics[width=0.30\textwidth]{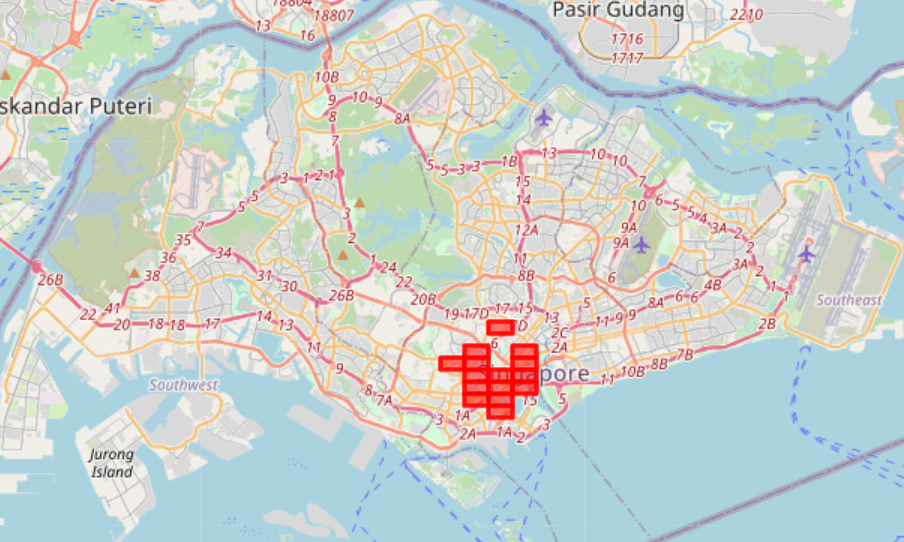}}\hfill
\subfloat[]{\includegraphics[width=0.32\textwidth]{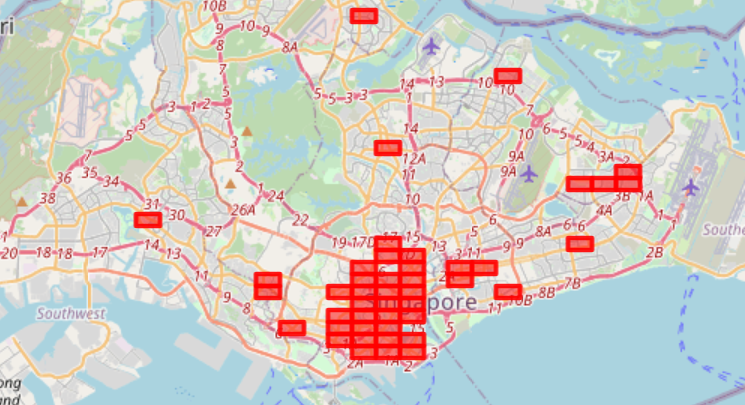}} \hfill
\subfloat[]{\includegraphics[width=0.32\textwidth]{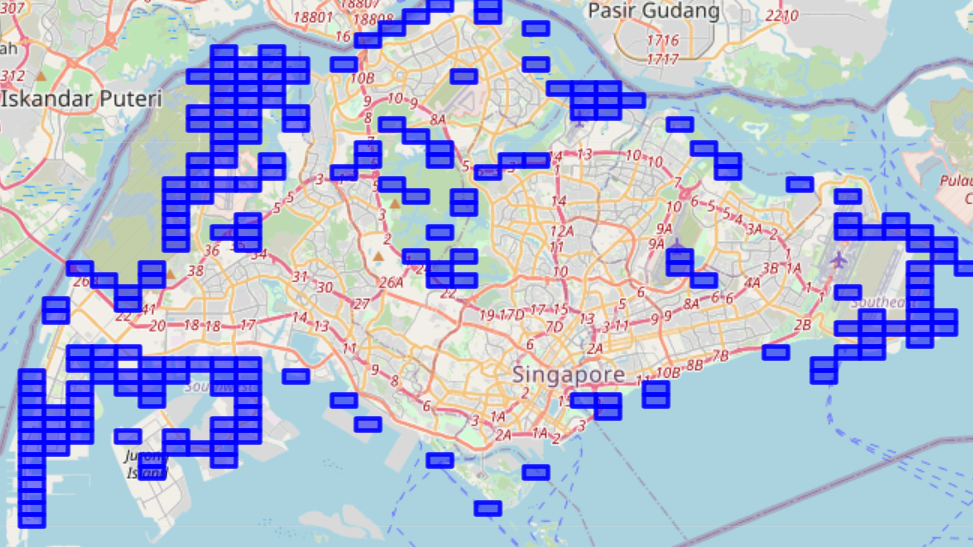}} 
\caption{\textit{Drivers' locations} with timely and late pickup.}%
\label{Fig:FastandSlowLocations}
\end{figure*}

\section{Co-clustering on driver-location-performance data}\label{sec:co-clustering}
In this experiment, we aim to extract some useful clusters and co-clusters from booking data to get driver-location specific insights. First, we prepare the rectangular relational data $D_{M \times N}$ where rows represent driver IDs, columns represent geohash grid IDs, and entries $D_{ij}$ represent the performance measure of a driver at a geohash grid, corresponding to the index $i$ and $j$. We considered late pickup rate (LPR) (\%) as a performance measure in our work. We normalized it to the interval $[0~1]$, where 0 indicates always timely (on and before time)  pickup and 1 indicates always late pickup. If a driver didn't have any booking at a location in our data, then we put $-1$ at the corresponding row and column in the relational matrix, to differentiate it with timely pickup (0\%). Co-clustering on this relational matrix would hopefully give us the following information: (i) driver cluster(s): group(s) of similar drivers e.g. drivers who were always on time or late to pick up passengers; (ii) location clusters(s): group(s) of similar driver locations e.g. locations where most drivers were always on time or late to pick up passengers; (iii) driver-location co-cluster(s): group(s) of locations where group(s) of drivers have same performance (e.g. late or on-time). 

\Fig~\ref{Fig:CoclusteringResults}(a) shows the color image of the relational data matrix, $I(D_{M \times N})$, for $M=48205$ drivers and $N= 845$ geohash grids, where blue indicates no bookings (-1), green indicates timely pickup (0), and red indicates late pickup (1). Between green to red spectrum (green<yellow<orange<red<bright red), the brighter the color, the later the driver pickup (or poorer pickup performance). Clearly, \Fig~\ref{Fig:CoclusteringResults}(a) does not show any noticeable cluster structure.

Next, we apply sco-iVAT algorithm on relational data $D_{M \times N}$, with the desired sampled size for MMRS sampling on row objects (driver IDs) as $m=1000$ and $k'= 100$. Since the number of columns (geohash grids) in $D_{M \times N}$ is not much large, we directly apply iVAT on column objects ($n=N$) without applying MMRS sampling as a prior step. 
 
The reordered relational matrix image (RRI) of $D_{M \times N}$ in~\Fig~\ref{Fig:CoclusteringResults}(b), $I({D}_{m \times n}^{*})$,  appears to show us some cluster structures on the right part of the image. We can see a vertical strip ($615 \times 30$) on the extreme right, shown with a red text arrow $1$. Most points within this cluster structure lie between yellow and red spectrum (from top to the bottom of the strip) indicating that the geohash grids, corresponding to the vertical strip ($1$), have higher late pickup rate for almost all drivers compared to other locations. We extracted the indices of these geohash grids from the reordered array of column objects $CP$, and show them in~\Fig~\ref{Fig:FastandSlowLocations}(a). We can see that these $30$ geohashes correspond to the busiest locations in Central Business District (CBD) area in Singapore such as Orchard Road, Bugis, Raffles Place, and City Hall that are mostly crowded due to many restaurants, shopping malls, tourist attractions, and big offices. Consequently, when drivers accept bookings at these locations, they mostly appear late for passenger pickups.

\Fig~\ref{Fig:CoclusteringResults}(b) also shows a horizontal strip ($45 \times 230$) at the bottom of the image (a large zoom may be required), as shown with a magenta text arrow $2$. Similar to the vertical strip, most of the datapoints in this cluster structure lie between yellow to red spectrum (from left to the right of the strip) indicating that these $45$ drivers are mostly late in their all bookings at these $230$ locations, corresponding to the horizontal co-cluster datapoints.~\Fig~\ref{Fig:FastandSlowLocations}(b) shows some of these geohash grids where these $45$ specific drivers were mostly late in their pickup bookings. These geohashes correspond to the CBD area, Harbourfront, Changi Airport, Woodland, Botanic Gardens and Jurong East. The analysis of these specific drivers is presented in penultimate paragraphs in this section.

Almost all the datapoints on the left side of the~\Fig~\ref{Fig:CoclusteringResults}(b), as shown in the vertical rectangle  (pointed out by arrow $3$), have blue pixels indicating that most drivers at these locations did not have any booking. ~\Fig~\ref{Fig:FastandSlowLocations}(c) shows these geohash grids conveying to us that these locations are in the outskirts of Singapore or are not accessible by road network. The remaining datapoints (no distinguishable cluster) in~\Fig~\ref{Fig:CoclusteringResults}(b) correspond to the timely pickup performance (in green and yellow) indicating that majority of the drivers complete their pickup on-time at the majority of the locations. From here on,  we call mostly late drivers as "late drivers" and remaining drivers as "on-time drivers" for conciseness.

\begin{figure}
\captionsetup[subfigure]{justification=centering}
\centering
\subfloat[Late pickup rate (\%) distribution for on-time and late drivers]{\includegraphics[width=0.25\textwidth]{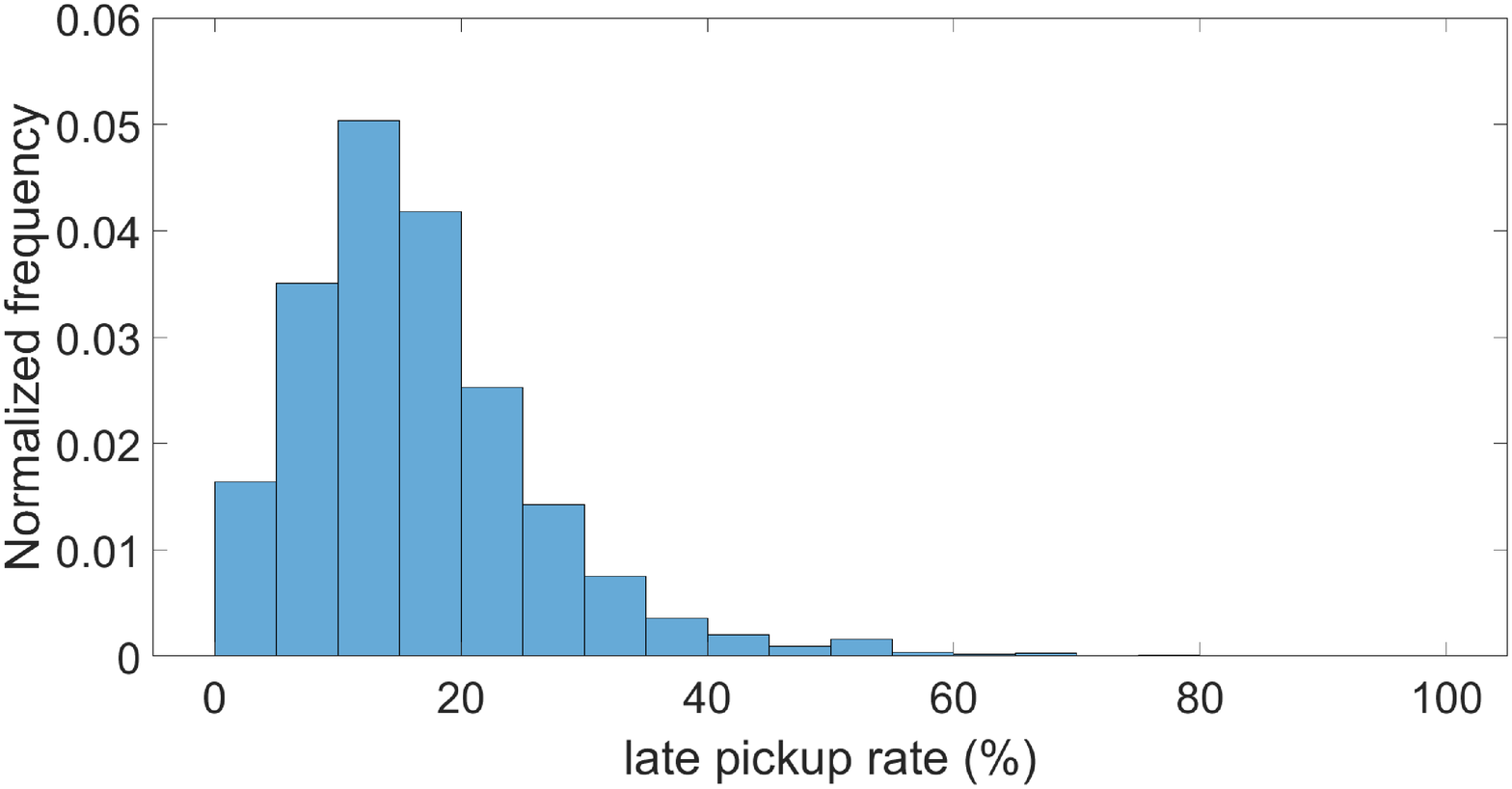}
\includegraphics[width=0.25\textwidth]{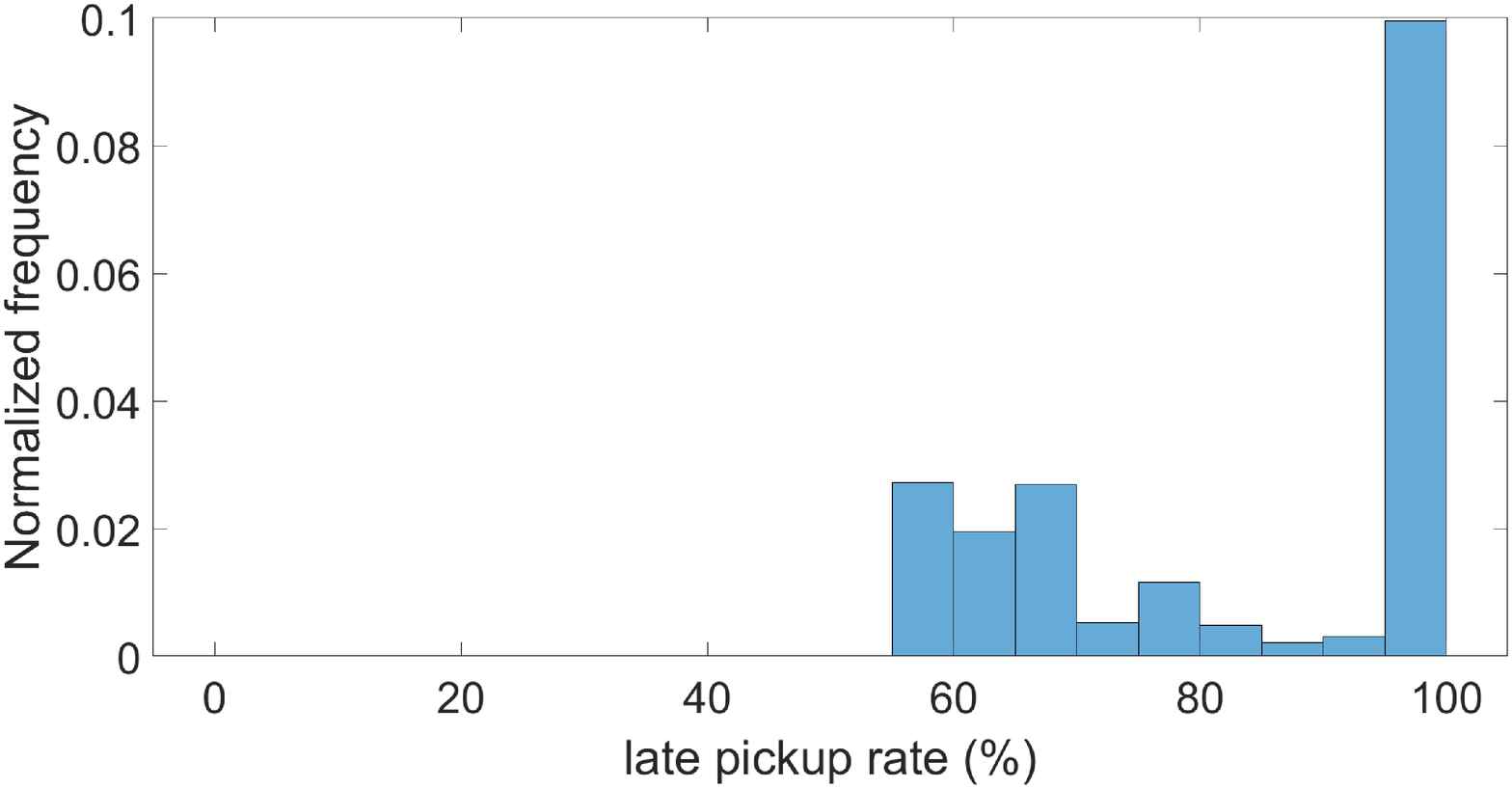}}\\
\subfloat[Total booking distribution for on-time and late drivers]{\includegraphics[width=0.25\textwidth]{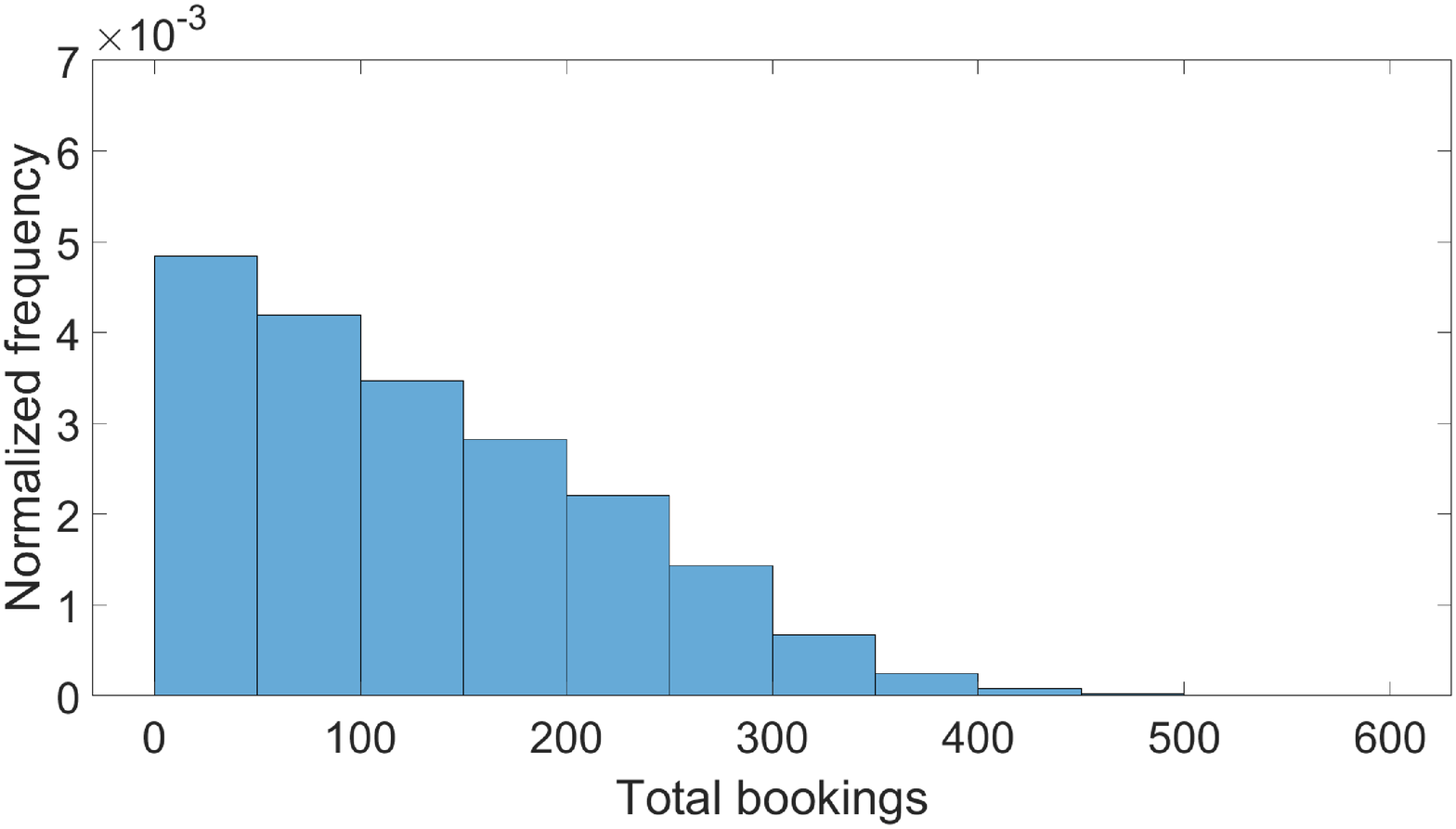}
\includegraphics[width=0.25\textwidth]{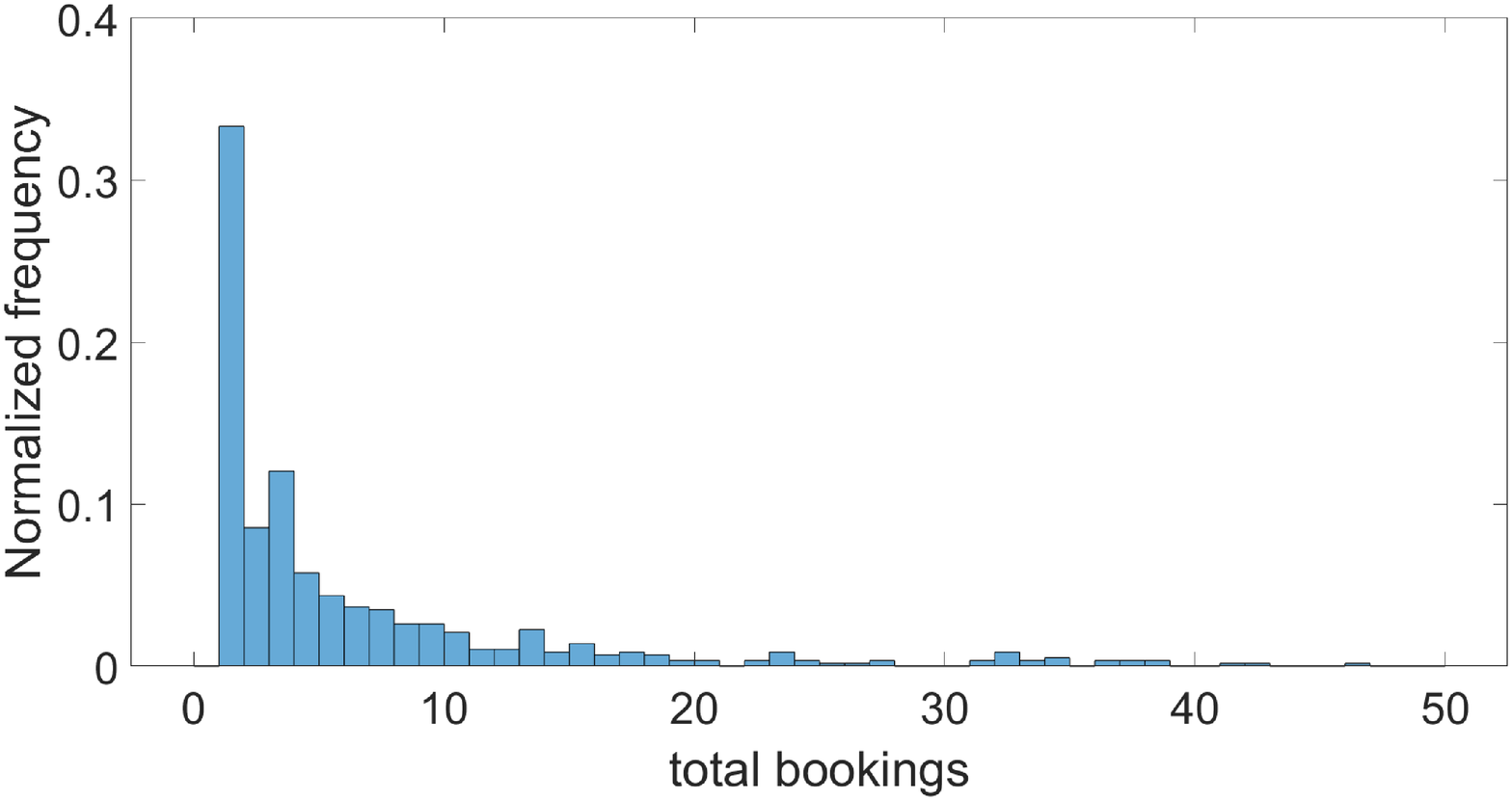}} 
\caption{Comparison of on-time and late drivers for passenger pickup.}%
\label{Fig:Normal_Late_Drives}
\end{figure}

We extracted the indices of the $45$ driver IDs from horizontal co-cluster $2$ (corresponding to the late pickups) and extended their labels to the remaining non-sampled driver IDs to identify similar drivers, using Step $9$ of the algorithm. After the extension, we identified a total of $308$ drivers who were mostly late for passenger pickups.~\Fig~\ref{Fig:Normal_Late_Drives} compares the $308$ late drivers (corresponding to co-cluster $2$) and  $47897$ on-time drivers based on the total booking and LPR (\%) distribution. It can be seen from~\Figs~\ref{Fig:Normal_Late_Drives} (a),(b)  that average LPR (\%) for on-time drivers is $10\%$ which indicates that most drivers were on-time for  $90\%$ of their total pickup bookings. In contrast, the late drivers had more than $60\%$ LPR.~\Figs~\ref{Fig:Normal_Late_Drives} (c) (d) show that most of the late drivers had less than $10$ bookings whereas the on-time drivers had on average $100$ bookings, thus reducing the average late pickup rate of on-time drivers by smoothing the effect of a few late pickups by many timely pickups. Note that the sco-iVAT algorithm was also applied on weekday/weekend and hourgroup based relational data, however, the results are not presented here due to space limitations.

\section{Timely Pickup Prediction}\label{sec:timelypickup}
Existing studies utilize the historical trajectory data to predict the ETA for a given trip or route. One of the main features of ride-hailing services is that a route is recommended to the driver to pick up the passenger. However, in practice, drivers may choose to follow a different route dynamically based on the traffic flow, conditions, or their prior knowledge about the locality. Then, a natural question would be: is it possible to predict the probability of timely passenger pickup for each driver candidate for a booking request, without using entire driver trajectories. Therefore, in this experiment, we explore the possibility of predicting the timely pickup of passenger for a given driver ID, \textit{driver location}, \textit{pickup location}, booking request timestamp and \textit{aggregated features} of drivers and locations from historical booking records. One may argue that due to apparent dynamicity in the traffic conditions, this goal can be challenging. However, our analysis and results in this experiment approve the possibility of such predictions that indirectly incorporates the drivers' familiarity about road-networks and their performance at different locations from \textit{aggregated features}. 

We considered the $1277732$ booking records and pre-computed aggregated features for classification, where each booking is represented by a $5$-tuple: (driver ID, \textit{driver location}, \textit{pickup location}, day of the week (weekday/weekend) and hourgroup). The Boolean variable is\_late\_pickup was considered as class label specifying timely (0) and late pickup (1) for each booking. We obtained a balanced subset (equal distribution in both class) based on the number of points in the minority class (class $1$) in the original data and then split it into training, test and validation set in $64:20:16$ ratio. We trained the \textit{logistic regression} (LR) model on training data for binary classification and evaluated it on testing and validation data based on the classification accuracy.


\begin{figure}
\centering
\includegraphics[width=0.48\textwidth, ]{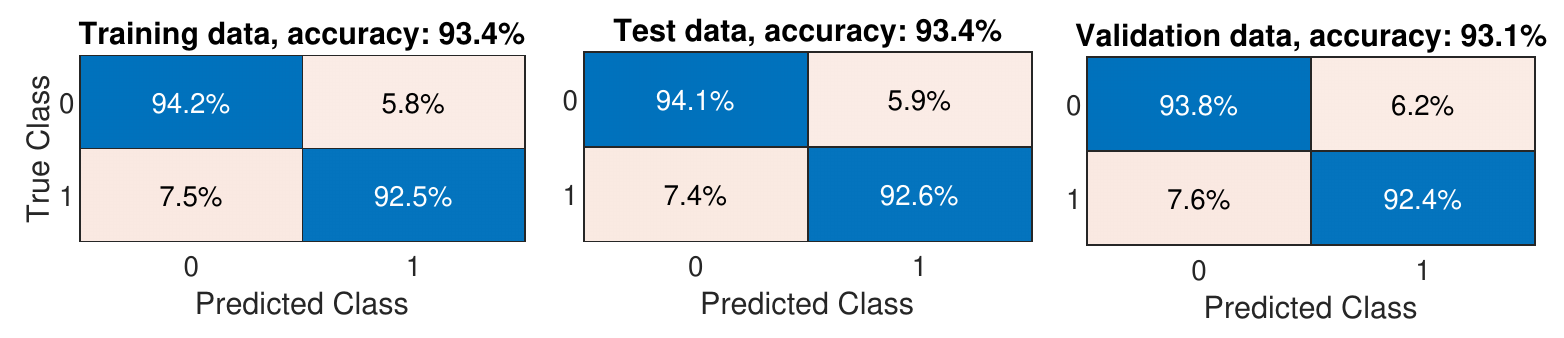}
\caption{Confusion matrix for timely pickup classification.}
\label{Fig:ConfusionMatrix}
\end{figure}

\begin{figure}
\centering
\includegraphics[width=0.5\textwidth]{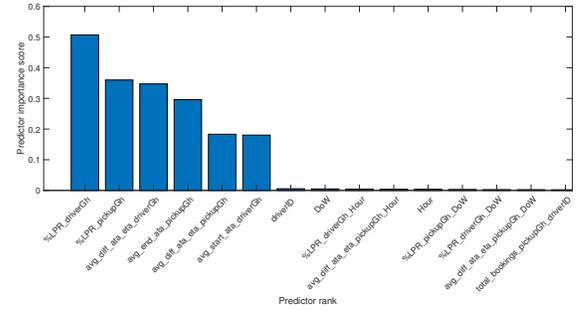}
\caption{Predictor importance for timely pickup prediction.}
\label{Fig:PredictorImportance}
\end{figure}

Fig~\ref{Fig:ConfusionMatrix} shows the confusion matrix for training, test and validation set. The LR model achieves $93\%$ classification accuracy for both test and validation set which indicates that it is possible to predict the timely pickup of the passenger for a driver at the time of booking request, even without using the entire trajectory data. We also study the importance of different predictors for timely pickup classification. In this regard, we utilize \textit{minimum redundancy maximum relevance} (mRmR)  algorithm that uses mutual information criteria between different features and response variable.~\Fig~\ref{Fig:PredictorImportance} shows the ranking of the top $15$ features based on their importance to the classification. Among them, \%LPR and avg\_diff\_ata\_eta at driver and pickup geohash, respectively, avg\_end\_ata, and avg\_start\_ata  turn out to be the six most importance predictor indicating that both \textit{driver location} (at the time of booking) and \textit{pickup location} significantly affects driver performance. 

\begin{figure}
\captionsetup[subfigure]{justification=centering}
\centering
\subfloat[avg\_start\_ata\_distribution]{\includegraphics[width=0.25\textwidth]{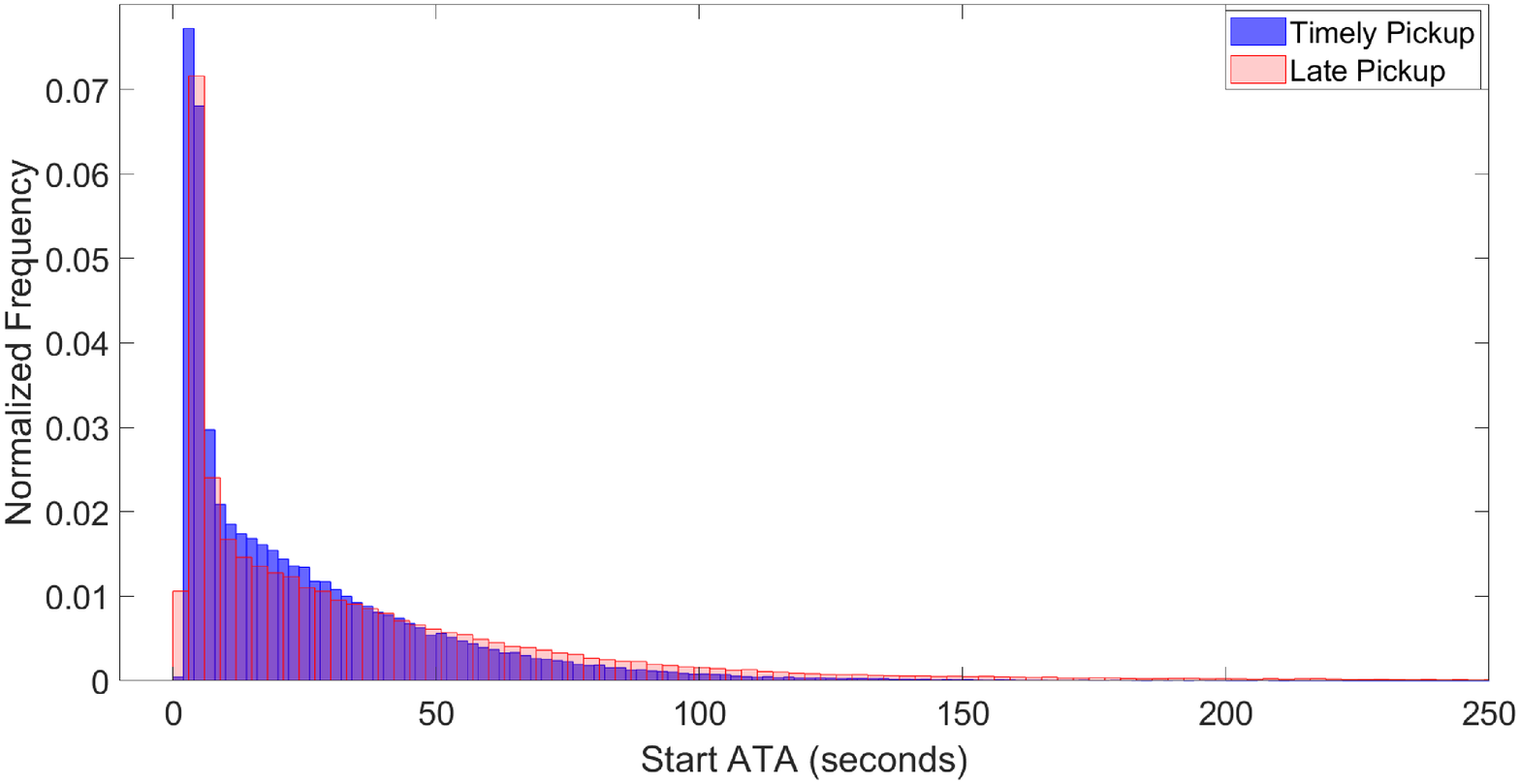}} 
\subfloat[avg\_end\_ata\_distribution]{\includegraphics[width=0.25\textwidth]{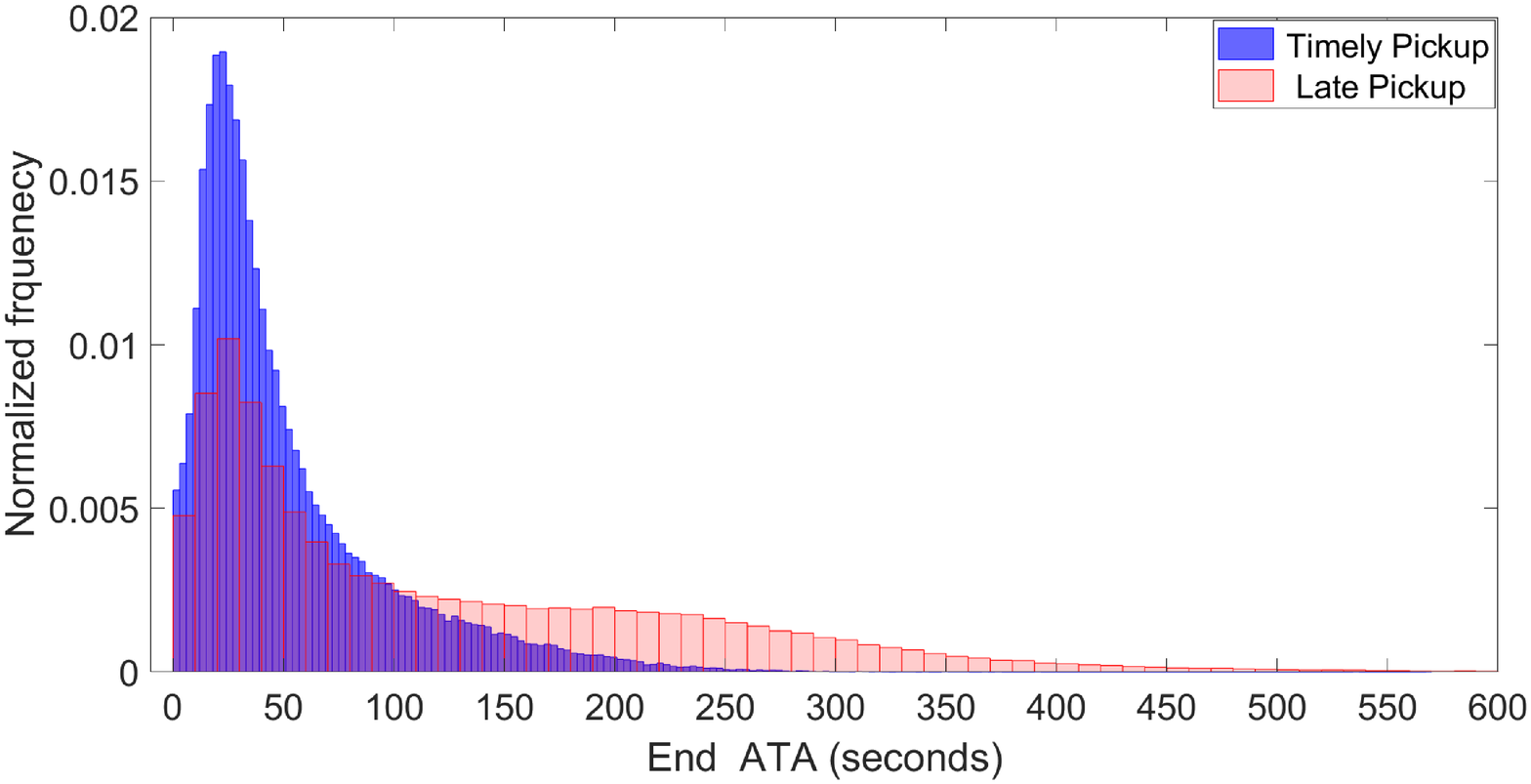}}
\caption{avg\_start\_ata (at \textit{driver locations}) and avg\_end\_ata (at \textit{pickup locations}) distribution.}%
\label{Fig:ATADistribution}
\end{figure}

We further analyze the avg\_end\_ata  and avg\_start\_ata features for timely pickup classification. \Fig~\ref{Fig:ATADistribution} shows the distribution of avg\_end\_ata  and avg\_start\_ata for timely and late pickup classes. It can be seen that the avg\_end\_ata is significantly higher for late pickups compared to the avg\_start\_ata which indicates that drivers usually take longer than the estimated time at pickup locations. This is probably because (i) most pickup locations in Singapore have designated pickup points, thus creating a long queue of taxis, especially near offices, tourist attraction, shopping malls etc., and (ii) drivers may take some time to find the passenger or a convenient pickup point at residential societies and colonies, especially when a driver is not much familiar to that area or locality. This also applies to the driver locations as, at the time of booking request, some drivers might be either at the crowded places, complex intersections or at unfamiliar places, thus ending up with spending more time before starting an easy and uninterrupted trip. Therefore, prior driving experience and familiarity about road network (e.g., driver and pickup location) should be taken account into the booking assignment framework in terms of driver's performance and total bookings at those locations.

\section{Driver scoring mechanism for booking assignment framework}\label{sec:scoring}
In this section, we devise two scoring mechanisms to compute the pickup performance score for each driver as a function of important aggregated features (mentioned in the previous section). In the first approach, we suggest a simple mechanism which computes the pickup performance score of a driver as a function of \% late pickup rate (LPR) of the driver, \textit{driver location} and the \textit{pickup location}, as given below:
\begin{multline}\label{Eq1}
    Score(driver\_ID, driverGh, pickupGh) =\\ \frac{1}{2}\left( \frac{{\%LPR}_{driverGh}}{{\%LPR}_{driver\_ID, driverGh}} +  \frac{{\%LPR}_{pickupGh}}{{\%LPR}_{driver\_ID, pickupGh}}\right),
\end{multline}
The score formulation in~\Eq~(\ref{Eq1}) indicates that drivers who have less \%LPR than the average \%LPR  (of all the drivers) at the corresponding driver and pickup geohash will achieve a higher score. This scoring mechanism is max-optimal i.e., the higher the score, the better the performance of the driver. 

As a second approach, we suggest a logistic regression based scoring mechanism that estimates the driver scores as an output probability for timely pickup (class $0$). In this mechanism, first the log-odds for class $0$ is computed as a linear combination of the important predictors and then the score (probability for class $0$) is computed, as given below:
\begin{eqnarray}
    \textrm{log-odds} =  (\alpha_{0}+\alpha_{1}f_{1}+\alpha_{2}f_{2}+....+\alpha_{p}f_{p}),\\
    Score = exp(\textrm{log-odds})/ (1+exp(\textrm{log-odds})),
\end{eqnarray} 
where $f_{1},f_{2},...,f_{p}$ are the important predictors such as \%LPR and avg\_diff\_ata\_ata at driver and pickup location, avg\_start\_ata, avg\_end\_ata, DoW etc., and $\alpha_{0},\alpha_{1},...,\alpha_{p}$ are bias term and corresponding feature weights which are learnt from regression model considering \%LPR of driver as an output score. The score in the second mechanism ranges in $[0~1]$, and higher score represents better performance of the driver. 

Note that once the aggregated features have been computed and weight parameters are learned offline, then scores in both mechanisms can be computed in real-time for all driver candidates at the time of booking request. While the first score mechanism is a simple formulation that considers only drivers' and locations' \%LPR,  the second scoring mechanism is a more advanced approach that considers all important predictors including their temporal distribution.

Ride-hailing service providers can incorporate these scores into their booking assignment model to prioritize\footnote{All drivers who completed less than a pre-defined number of bookings (say $5$) at the corresponding location, should be given the same score as the top drivers so that newly joined drivers are not penalized from our scoring mechanisms.} well-performing drivers for a booking request.  Numerical evaluation and validation of both mechanisms using A/B testing, for booking assignment model are beyond the scope of this paper, and we intend to pursue it in our future work. 

\section{Conclusion}\label{sec:conclusion}
This article presents both exploratory and confirmatory analysis of taxi drivers' locations at the time of booking request and pickup locations for drivers' pickup performance. To facilitate exploratory analysis, we implemented a modified and extended version of a co-clustering technique, called sco-iVAT, to obtain useful clusters and co-clusters from a big relational data. We applied sco-iVAT on a relational data matrix, derived from the booking data of Grab ride-hailing service, and identified useful co-clusters among drivers and their locations at the time of booking request, based on their pickup performance.

In the confirmatory analysis, we explored the possibility of predicting timely pickup for a driver given a booking request. 
 We extracted several important features from historical booking records based on the spatio-temporal activities of drivers for passenger pickups. The high classification accuracy ($93\%$) on Grab data suggests that timely passenger pickup is predictable for a driver, even without utilizing full trajectory data. We also devised two scoring mechanisms to compute the pickup performance score for each driver candidates for a booking request that could be integrated by e-hailing service providers in their booking-assignment model to prioritize good-performing drivers. In our future work, we aim to implement driver scoring mechanism on a real platform, and evaluated their effects for pickup performance.  

\section{Acknowledgment}
This work was funded by the Grab-NUS AI Lab, a joint collaboration between GrabTaxi Holdings Pte. Ltd. and National University of Singapore, Singapore.
\bibliographystyle{IEEEtran}
\bibliography{Bibliography}
\vskip -2\baselineskip plus -1fil
\begin{IEEEbiography}[{\includegraphics[width=1in,height=1.25in,clip,keepaspectratio]{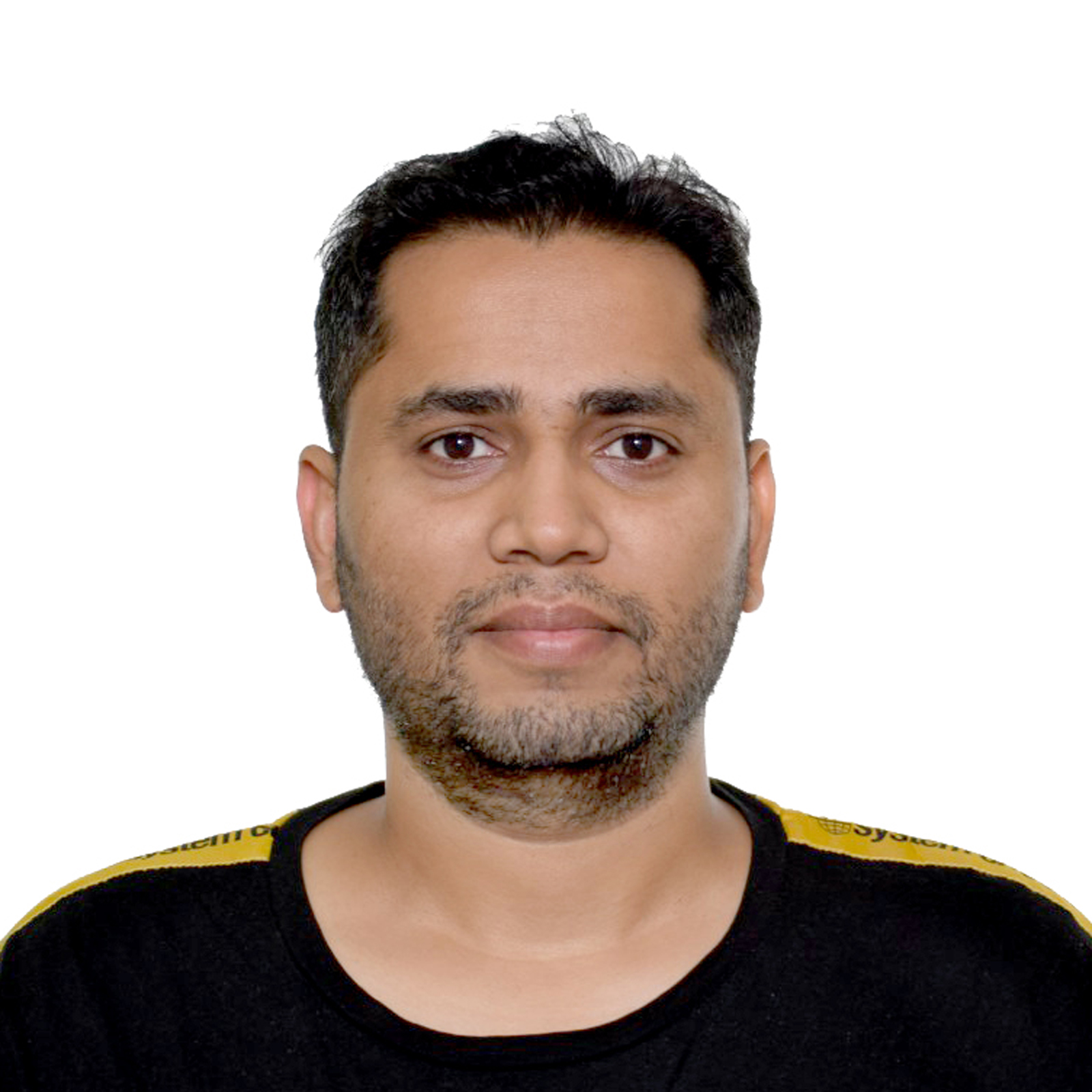}}]%
{Punit Rathore} is a Postdoctoral Research Fellow in Senseable City Lab at Massachusetts Institute of Technology (MIT) Cambridge, Previously, he worked as a Postdoctoral Researcher at Institute of Data Science, National University of Singapore (NUS) Singapore. He received the M.Tech degree in Instrumentation Engineering from Indian Institute of Technology, Kharagpur, India in 2011, and Ph.D. degree from the University of Melbourne, Melbourne, Australia in 2019. His research interests include big data cluster analysis, anomaly detection, urban data analytics, and Internet of Things.
\end{IEEEbiography}
\vskip -2\baselineskip plus -1fil
\begin{IEEEbiography}[{\includegraphics[width=1in,height=1.2in,clip,keepaspectratio]{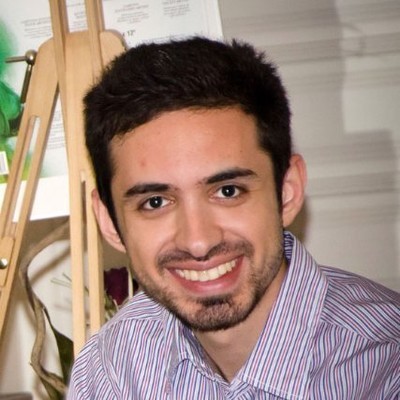}}]%
{Ali Zonoozi} received his B.Sc. in Computer Engineering in 2010 and M.Sc. in Software Engineering in 2013 from Amirkabir University of Technology, Tehran, Iran. He completed his joint Ph.D. from Nanyang Technological University and Agency for science, Technology and Research (A*STAR), Singapore in 2018. Current, he is a Lead Data Scientist at Grabtaxi Ltd. Singapore since 2018 where he has been working on map improvement using Computer Vision techniques, Allocation optimization, prioritisation logics, and real-time NLP services.
\end{IEEEbiography}
\vskip -2\baselineskip plus -1fil
\begin{IEEEbiography}[{\includegraphics[width=1in,height=1.2in,clip,keepaspectratio]{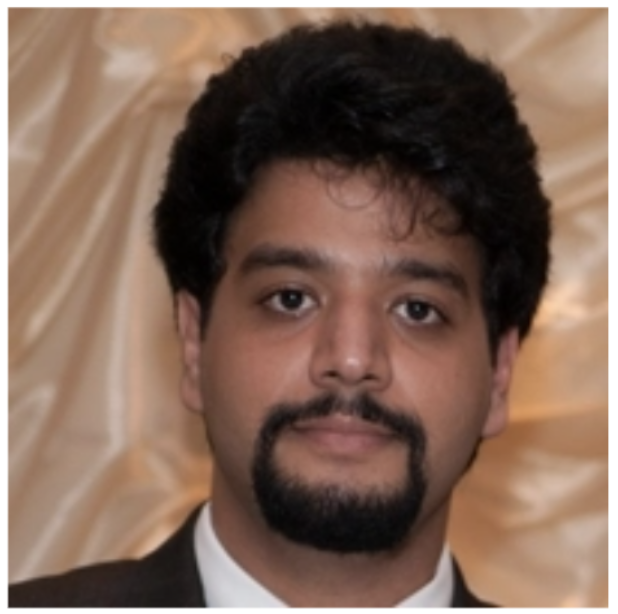}}]%
{Omid Geramifard} received B.Sc. in Computer Engineering from Isfahan University of Technology, Isfahan, Iran, in 2008, and Ph.D. degree in Machine Learning from NUS, Singapore, in 2013.  After his Ph.D., he worked at Singapore Institute of Manufacturing Technology as a Research Scientist (2013-17) and at Grab as a Senior and Lead Data Scientist positions working on allocation problems, allocation overseeing the prioritisation logics, allocation strategies and algorithms during 2017-2019. He is currently the head of Airline Data in AirAsia Group overseeing all the Airline and Operations data related projects.
\end{IEEEbiography}
\vskip -2\baselineskip plus -1fil
\begin{IEEEbiography}[{\includegraphics[width=1in,height=1.2in,clip,keepaspectratio]{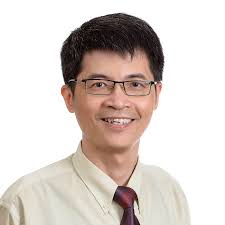}}]%
{Kian-Lee Tan}  received the PhD degree in Computer Science from NUS in 1994. He is a Professor with
the School of Computing at NUS. His current research interests include query processing and optimization in multiprocessor and distributed systems, database performance, data analytics, and database security. 
He was a co-recipient of Singapore's President
Science Award in 2011. He is also a 2013 IEEE
Technical Achievement Award recipient. He is an
Associate Editor of the ACM Transactions on Database Systems (TODS) and the World Wide Web Journal. He has also served on the editorial boards of the VLDB Journal and the IEEE Transactions on Knowledge and Data Engineering (2009-2013).
\end{IEEEbiography}
\end{document}